\documentclass[fleqn,usenatbib]{mnras}

\usepackage{newtxtext,newtxmath}

\usepackage[T1]{fontenc}
\usepackage{ae,aecompl}

\usepackage{graphicx}	
\usepackage{amsmath}	
\usepackage{amssymb}	
\usepackage{threeparttable}

\title[Deep GMRT 610 MHz Observations of the ELAIS N1 Field]{Deep GMRT 610 MHz Observations of the ELAIS N1 Field : Catalogue and Source Counts}

\author[Ocran et al.]{
E.\ F. Ocran$^{1,3}$\thanks{E-mail: ocran62@gmail.com}, A.\ R.\ Taylor$^{1,2,3}$, M.\ Vaccari$^{2,3,4}$, 
C.H. Ishwara-Chandra$^{3,5}$,\and I.\  Prandoni$ ^{4}$
\\
$^1$ Department of Astronomy, University of Cape Town, Private Bag X3, Rondebosch 7701, South Africa \\
$^2$ Department of Physics and Astronomy, University of the Western Cape, Private Bag X17, Bellville 7535, South Africa \\
$^3$ Inter-University Institute for Data Intensive Astronomy, South Africa \\
$^4$ INAF - Istituto di Radioastronomia, via Gobetti 101, 40129 Bologna, Italy \\
$^5$National Centre for Radio Astrophysics, Tata Institute of Fundamental Research, Pune 411007, India\\
}

\date{Accepted 2019 October 17. Received 2019 October 17; in original form 2019 June 23}

\pubyear{2019}

\usepackage{etoolbox}
\makeatletter
\patchcmd\@combinedblfloats{\box\@outputbox}{\unvbox\@outputbox}{}{\errmessage{\noexpand patch failed}}
\makeatother
\begin{document}
\label{firstpage}
\pagerange{\pageref{firstpage}--\pageref{lastpage}}
\maketitle

\parskip0pt
\baselineskip12pt

\begin{abstract}
This is the first of a series of papers based on sensitive 610 MHz observations of the ELAIS N1 field using the Giant Metrewave Radio Telescope.  
We describe the observations, processing and source catalogue extraction from 
a deep image with area of 1.86\,deg$^{2}$ and minimum noise of $\sim$7.1\,$\mu$Jy/beam.
We compile a catalogue of 4290 sources with flux densities in the range 28.9\,$\mu$Jy - 0.503\,Jy, and derive the Euclidean-normalized differential source counts for sources with flux densities brighter than $\rm{35.5\,\mu\,Jy}$.
Our counts show a flattening at 610 MHz flux densities below 1 mJy. 
Below the break the counts are higher than previous observations at this frequency, but generally consistent with recent models of the low-frequency source population.
The radio catalogue is cross-matched against multi-wavelength data leading to identifications for 92\% and reliable redshifts for 72\% of our sample, with 19\% of the redshifts based on spectroscopy. 
For the sources with redshifts we use radio and X-ray luminosity, optical spectroscopy and mid-infrared colours to search for evidence of the presence of an Active Galactic Nucleus (AGN).
We compare our identifications to predictions of the flux density distributions of star forming galaxies (SFGs) and AGN, and find a good agreement assuming the majority of the sources without redshifts are SFGs. 
We derive spectral index distributions for a sub-sample. The majority of the sources are steep spectra, with a median spectral index that steepens
with frequency; $\mathrm{\alpha^{325}_{610}\,=\,-0.80\,\pm\,0.29}$, $\mathrm{\alpha^{610}_{1400}\,=\,-0.83\,\pm\,0.31}$ and $\mathrm{\alpha^{610}_{5000}\,=\,-1.12\,\pm\,0.15}$.
\end{abstract}



\section{Introduction}

The study of the faint radio continuum universe and of its properties has recently become a very active field of research not only because of the planned transformational capabilities of the Square Kilometre Array \citep{2015aska.confE.174B} on this field, but also because of the major steps being taken and planned with SKA pathfinders and precursors. 
Deep radio observations of the extragalactic sky are a powerful means to probe the properties of diverse source populations over a variety of environments to high redshift \citep{1984ApJ...287..461C,1995ApJ...450..559B,1999MNRAS.305..297G}.
Radio emission is important for galaxy population studies, as the synchrotron emission is a clear indicator of activity for both star formation galaxies (SFGs) and active galactic nuclei (AGN). 
Moreover, radio emission is not affected by dust obscuration, hence can probe astrophysical processes to large distances.
 
At faint radio flux densities star forming galaxies dominate. These are very different  from the radio sources seen in the bright radio sky \citep{2011ApJS..193...27W,2016A&ARv..24...13P}, which are dominated by Active Galaxies. Counts of radio galaxies versus flux density provide useful information, as the source count shape is directly related to the evolutionary properties of the galaxies \citep{2001A&A...365..392P,2010A&ARv..18....1D,2011ApJ...740...20P,2015MNRAS.452.1263P}. 
Radio source counts represent the most immediate observational constraint to evolutionary models 
of radio sources (\citet{2001A&A...365..392P}.
The now well-established flattening of the counts below 1 mJy is interpreted as the signature 
of  the rise of star-forming galaxies \citep{2004MNRAS.355L...9R,2008MNRAS.386.1695S,2009ApJ...694..235P}.

Surveys at low frequencies (e.g. \citealt{2008MNRAS38375G, 2009MNRAS.395..269S,2016MNRAS.460.2385W}) are an important complement to higher frequency observations.
Low-frequency observations are powerful at detecting
ultra-steep spectrum sources, which are often galaxies at high redshifts (e.g. \citealt{1998MNRAS.301L..15B,2003MNRAS.346..627B,2008A&ARv..15...67M}). 
Combining low- and high-frequency radio observations allows studies of the radio continuum
spectra (e.g. \citealt{doi:10.1093/mnras/stw2638,2016MNRAS.463.2997M}), providng a more precise
characterization of the source properties \citep{2016MNRAS.462..917R}. 
 Radio spectral indices can be used to identify GHz peaked sources (GPS) (\citealt{1998JApA...19...63A,1998PASP..110..493O,2000MNRAS.319..429S}), ultra-steep spectrum sources (USS)  (\citealt{1994A&AS..108...79R,2001MNRAS.327..907J}), and core-dominated radio-quiet AGN \citep{2007ApJ...668L.103B}.

Star-forming galaxies are observed to have a mean spectral index between -0.8 and -0.7 at 1.4 GHz 
($S(\nu) \propto \nu^{\alpha}$),
with a relatively small dispersion of $\rm{\pm0.24}$ \citep{1992ARA&A..30..575C}. 
Studies combining 610-MHz and 1.4-GHz data, have found evidence for flatter spectral indices \citealt{2007A&A...463..519B,2008MNRAS38375G} and larger dispersions at sub-mJy radio flux densities \citep{2008MNRAS.383..479M}, suggesting that core-dominated radio-quiet AGN are playing a key role in the sub-mJy radio population.
\cite{2009MNRAS.397..281I} reported statistical analyses showing no clear evolution for the median spectral index, $\rm{\alpha^{610}_{1400}}$, as a function of flux density based on observations of the Lockman Hole using the Giant Metrewave Radio Telescope (GMRT). 
Their study found $\rm{\alpha^{610}_{1400}}$ to be $\rm{-0.6}$ to $\rm{-0.7}$. They also analyse the spectral indices based independently on GMRT and VLA selected samples, and found that a 610 MHz selected catalogue naturally tends to prefer the detection of steep-spectrum sources while selection at 1.4 GHz favours flatter spectra.

The European Large Area ISO Survey (ELAIS) N1 field has been observed at multiple radio frequencies. 
\citet{2009MNRAS.395..269S} observed it at 325 MHz using the Giant Metrewave Radio Telescope (GMRT), with the objective of identifying active galactic nuclei and starburst galaxies and examining their evolution with cosmic epoch.
\cite{2010ApJ...714.1689G} observed 15\,deg$^2$ with the Dominion Radio Astrophysical Observatory
synthesis telescope at 1420\,MHz to a minimum rms of 55\,$\mu$Jy\,beam$^{-1}$ in Stokes I and 45\,$\mu$Jy\,beam$^{-1}$ in polarisation.
\citet{2011ApJ...733...69B} observed 10\,deg$^2$ at 1.4 GHz with the JVLA in B configuration to a minimum rms of 87\,$\mu$Jy\,beam$^{-1}$ in total intensity.
\cite{2014ASInC..13...99T} observed a smaller region (0.13\,deg$^2$) at 5 GHz with the JVLA in B and C configuration to a minimum rms of 1.05\,$\mu$Jy\,beam$^{-1}$.


In this paper we present deep GMRT observations at 610 MHz of the ELAIS N1 field (EN1) covering 1.86 deg$^2$. The EN1 field is a large northern field that has been targeted by surveys spanning the entire electromagnetic spectrum. Further building on the extensive radio coverage of ELAIS\,N1 and wealth of multi-wavelength observations that provides valuable insights into galaxy formation and evolution. In this work, we achieve a minimum noise of 7.1 $\mu$Jy\,beam$^{-1}$ and an angular resolution of {\bf $\rm{6\,arcse \times\,6\,arcsec }$}.

The remainder of this paper is divided as follows: we first introduce the observations and data processing in Section~\ref{dataradio.sec}. In Section~\ref{sourcecounts.sec}, we present the 610-MHz source counts analysis. Section~\ref{crossmatch.sec} provides the multi-wavelength cross-identifications and the nature of the source population. The multi-frequency spectral analysis of the sample is presented in Section~\ref{spectral.sec}.
In this paper we assume a flat cold dark matter ($\rm{\Lambda}$CDM) cosmology with  $\rm{\Omega_{\Lambda} \ = \ 0.7}$, $\rm{\Omega_{m} \ = \ 0.3}$ and $\rm{H_{o} \ = \ 70 \ km\,s^{-1} \ Mpc^{-1}}$ .

\section{Observations and data processing}\label{dataradio.sec}

\subsection{Radio data}

The ELAIS N1 field was originally chosen for deep extragalactic observations with the Infrared Space Observatory (ISO) due to its low infrared background \citep{RowanRobinson2004,Vaccari2005}. Since then, it has  become one of the best-studied 1-10 deg$^2$ extragalactic fields. GMRT observations of the ELAIS N1 field were obtained during several observing runs from 2011 to 2013. The observations were carried out for 7 positions arranged in a hexagonal pattern centred on 
$\mathrm{\alpha \ = \ 16^{h} \ 10^{m} \ 30^{s}, \  \delta \ = \ 54^{\circ} \ 35 \ 00^{\prime \prime}}$ (see \citealt{2017MNRAS.468.1156O}). In this paper we present a deeper radio image, an improved radio data reduction and multi-wavelength analysis of the ELAIS N1 610 MHz Deep Survey first described by \cite{2017MNRAS.468.1156O}. 

The survey consists of 7 closely-spaced GMRT pointings, with on source integration time of $\sim$ 18 hours per pointing. 
The pipeline was modified to restrict the flags, which resulted in slightly less data being flagged. The shallower pointings were added at the edges where the deep and shallow pointing had rms within a factor of two.  The weights used in the mosaic follows the same procedure as in \cite{2017A&A...598A..78I}, Section 3.3. The weight is the inverse square of the local background rms noise (the inverse variance).  The resolutions before mosaic, for each pointing were in the range 4.5 to 6 arcseconds.
To reduce the noise around the edges of the mosaic image we included data from a set of pointings with 3 hours of observation each that are part of a wider but shallower study of ELAIS\,N1 (\citealt{2019MNRAS_Ishwara-Chandra}, in prep).

The data was analysed using CASA (Common Astronomy Software Applications) using standard procedures. The flux density scale was set using the primary calibrators 3C286 and 3C48, which were observed both at the start and at the end of each observing session. A phase calibrator was observed for 5 minutes every 30 minute of  target observations for phase and gain calibrations. After initial flagging using {\tt flagdata}, delay, bandpass and gain calibration was carried out. Post-calibration, the data was flagged again and re-calibrated. 
Channel averaging was done with post-averaging channel width of 0.78 MHz in order to keep the bandwidth smearing negligible. Split files from each pointing from different observing runs were combined using {\tt concat} before imaging. We used  {\tt tclean} for imaging.  Four rounds of phase-only self-cal and then 5 rounds of amplitude and phase self-cal was carried out on each pointing. The rms noise on the individual images were $\sim$ 15 $\mu$Jy/beam before mosaicing.  The primary beam correction and mosaic was carried out in AIPS using the python script \textit{make mosaic} (Intema, private communication) using a circular restoring beam of 6 arcsec.

\begin{figure*}
\centering
\centerline{\includegraphics[width=0.98\textwidth]{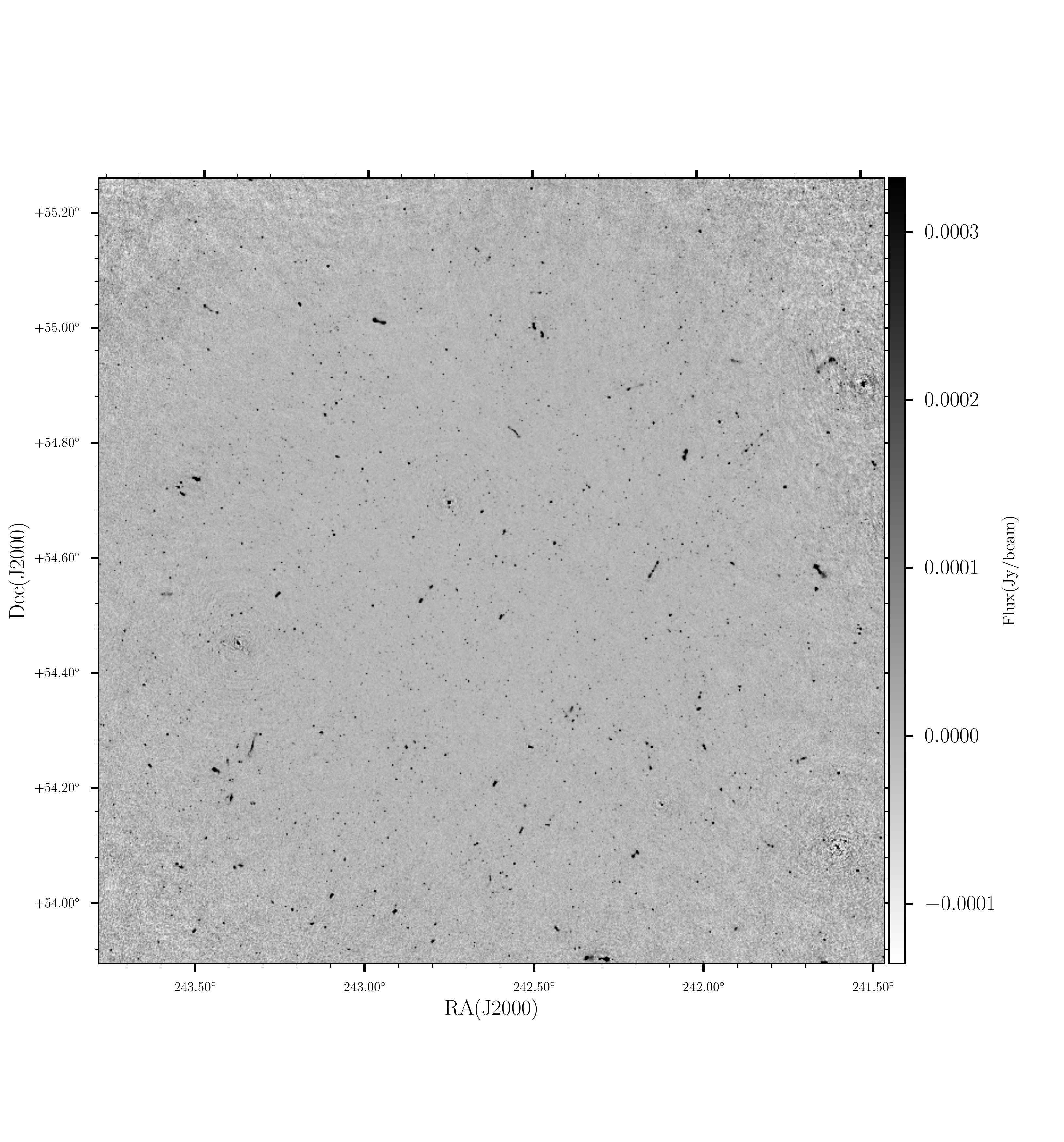}}
\caption{Image of the ELAIS\,N1 GMRT 610 MHz mosaic centered at $\mathrm{\alpha \ = \ 16^{h} \ 10^{m} \ 30^{s} , \  \delta \ = \ 54^{\circ} \ 35 \ 00^{\prime \prime}}$ (J2000). This image is 47$'$ on each side. The restoring beam is 6 arcsec circular and the RMS in the central region is $\sim$ 7.1\,$\mu$Jy\,beam$^{-1}$. The grey-scale brightness stretch ranges between -0.1 and 0.3 mJy beam$^{-1}$.} 
\label{fig:map} 
\end{figure*}

\begin{figure}
\centering
\includegraphics[width = 0.52\textwidth]{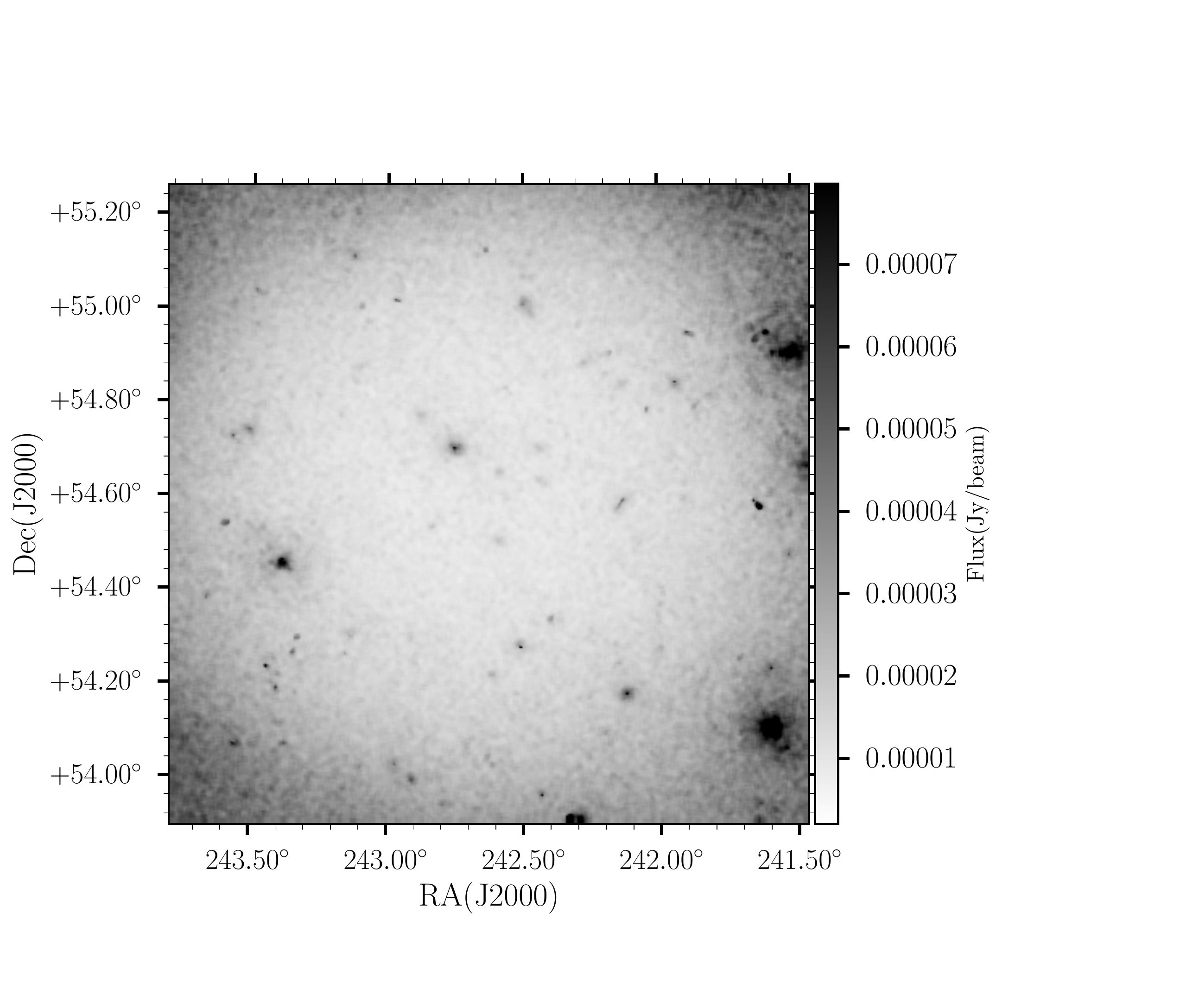}
\caption{Greyscale image showing the local rms noise of the final mosaicked GMRT image, derived using \textsc{PyBDSF} (see Section~\ref{Source finding}).  The grey-scale brightness stretch ranges between 0.01 and 0.07 mJy beam$^{-1}$.}
\label{fig:rms_area} 
\end{figure}

\begin{figure}
\centering
\includegraphics[width = 0.5\textwidth]{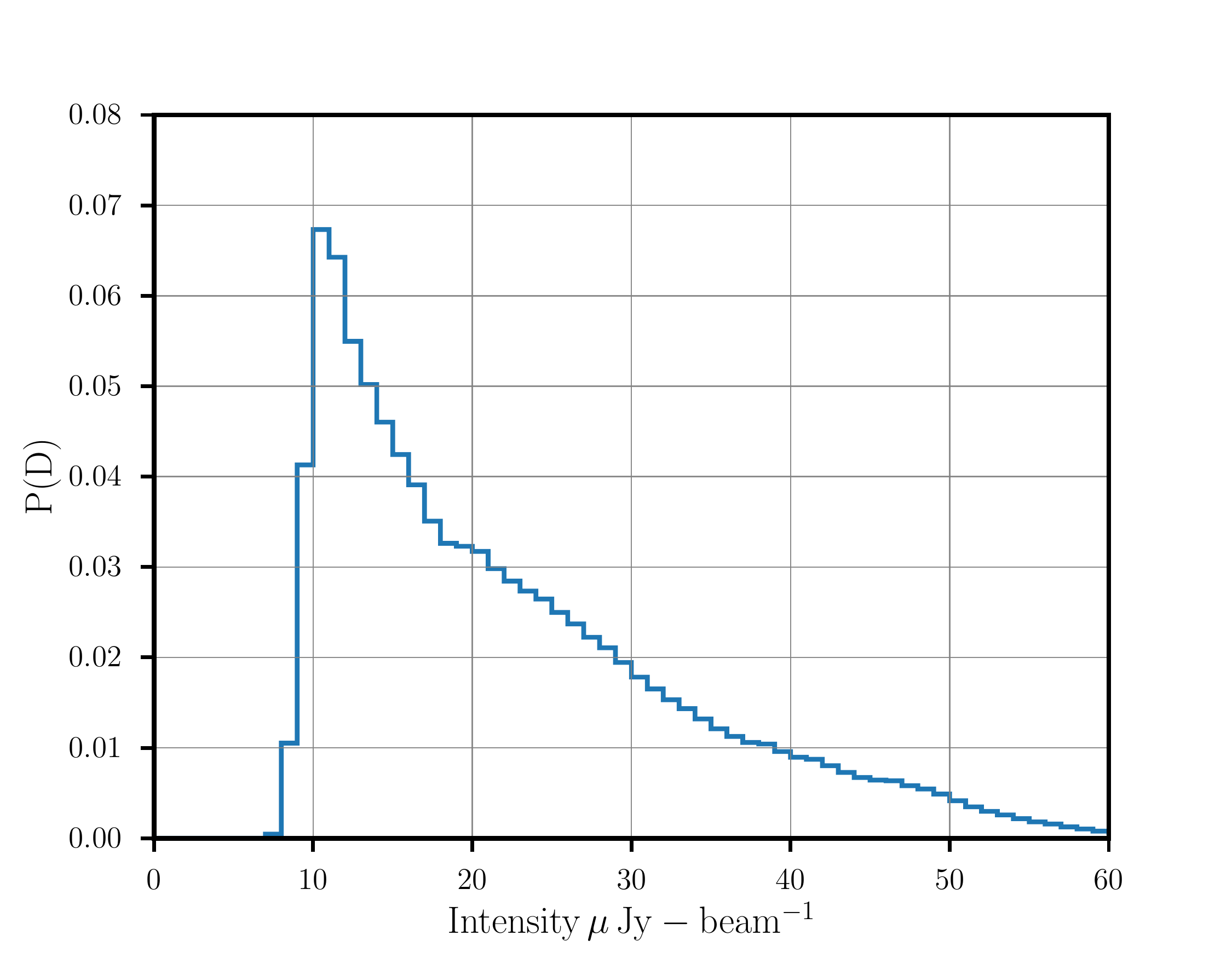}
\caption{The distribution of the 610-MHz rms for the GMRT sample. The mean and median rms are 22.70\,$\mu$Jy and 19.50\,$\mu$Jy respectively.}
\label{fig:rms_dist} 
\end{figure}

An image of the mosaic is shown in Figure~\ref{fig:map}. 
There are a small number of bright classical radio galaxies with double-lobed and jet morphologies, however most of the sources are compact as typically found in faint (sub-) mJy radio fluxes.
Figure~\ref{fig:rms_area} shows an image of the rms map created by \textsc{PyBDSF}, and Figure~\ref{fig:rms_dist} shows the distribution of pixel amplitudes in the rms image. 
The minimum rms noise in the central region of the image is 7.1 $\mu$Jy\,beam$^{-1}$.   The median noise in the mosaic is 19.5$\mu$Jy\,beam$^{-1}$.  The higher noise values arise primarily due to enhanced rms in small regions around very bright sources and from the lower mosaic weights at the edge 
of the mosaic. 

\subsection{Source Finding and Cataloguing}\label{Source finding}

The catalogue of radio sources was extracted using the \textsc{PyBDSF} source finder \citep{2015ascl.soft02007M}.
The rms map was determined with a sliding box \texttt{rms\_box}$\rm{\,=\,(80,\,10)}$ pixels (i.e. a box size of 80 pixels every 10 pixels), with a smaller box
\texttt{rms\_box\_bright}$\rm{\,=\,(40,\, 5)}$ pixels in the regions around bright sources to account for the increase in local rms as a result of calibration artefacts. 
Figure~\ref{fig:rms_area} illustrates the variation in rms noise determined across the entire mosaic image.

\textsc{PyBDSF} extracts sources by first identifying islands of contiguous emission above a given threshold \texttt{thresh\_isl}$\rm{=3\sigma}$, around pixels brighter than a given flux \texttt{thresh\_pix}$\rm{=5\sigma}$. Then it 
decomposes the islands into Gaussian components. It then combines significantly overlapping Gaussians into sources, and determines the flux densities, shapes and positions of sources \citep{2011A&A...535A..38I}. We used the \texttt{group\_tol} parameter with a value of 10.0 to allow more Gaussians to be grouped together and larger sources to be formed. 
Sources are classified as $'S'$ for single sources and $'M'$ for multiple Gaussian sources. The total number of sources detected by \textsc{PyBDSF} in the image is 6605 comprising of 7919 Gaussian components of which 5682 were single component sources. 

The catalogue consists of 4303 radio sources with signal to noise cut ($\rm{SNR\,=\,\frac{S_p}{rms}}$) above the $5\sigma$ threshold and flux threshold cut $\rm{\ge\, 0.1 \times RMS_{median}}$. 128 of the sources included in the catalogue flagged as having poor Gaussian fits. In this case the integrated flux density is the total flux measured in the island instead of that defined by the Gaussian fit.
\subsection{Multiple component sources}\label{Multi_component}
In order to generate a final source catalogue we need to identify multi-component sources that have not been recognised as such by \texttt{PyBDSF}, and therefore appear as a distinct radio source in the catalogue. This can happen when there is no significant radio emission between two radio lobes, or the local rms noise is overestimated because of large-scale faint radio emission (see \citealt{2017A&A...602A...1S}), which affects the ability of PyBDSF to properly detect the source.

\begin{figure}
 \centering
 \centerline{\includegraphics[width = 0.56\textwidth]{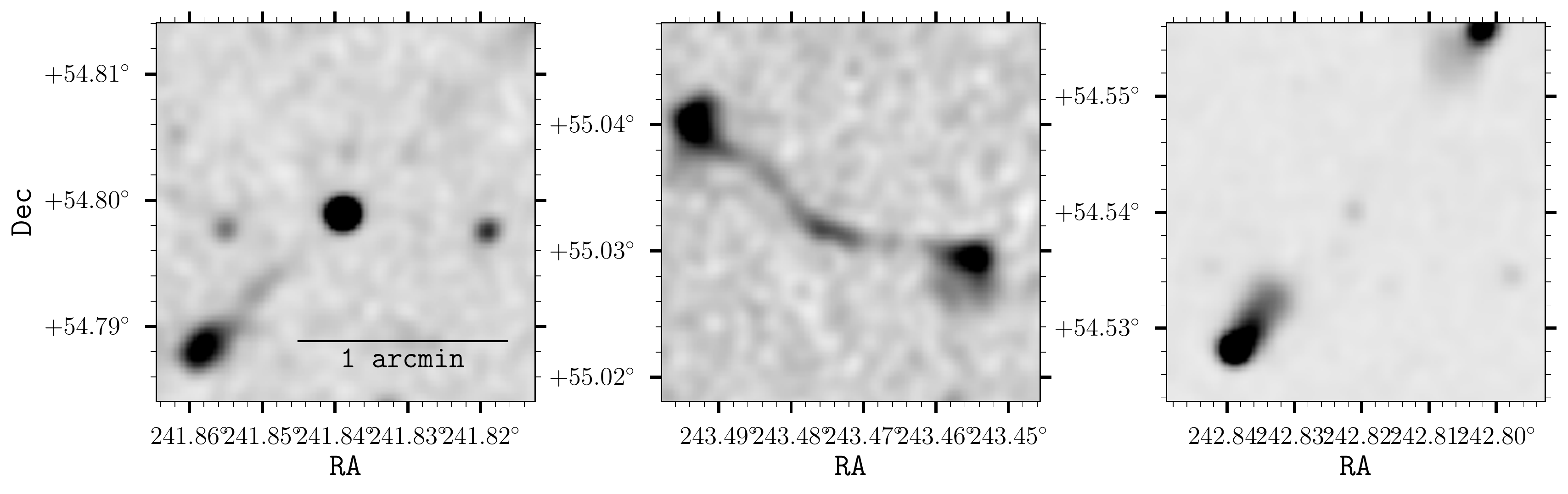}}
 \centerline{\includegraphics[width = 0.56\textwidth]{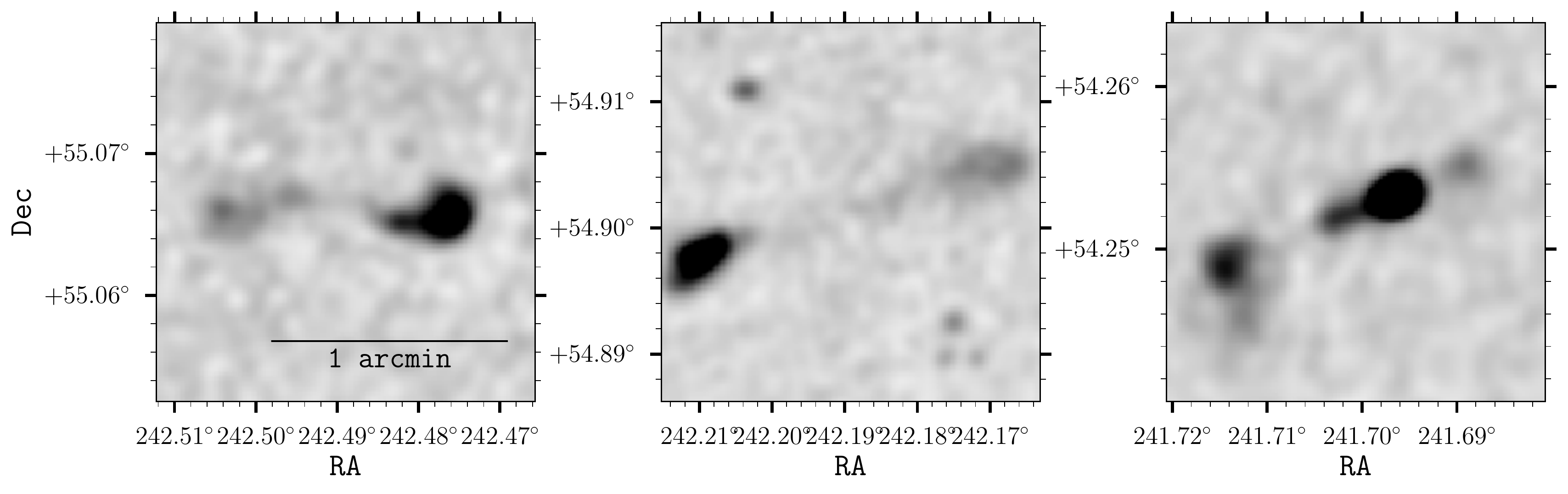}}
 \centerline{\includegraphics[width = 0.56\textwidth]{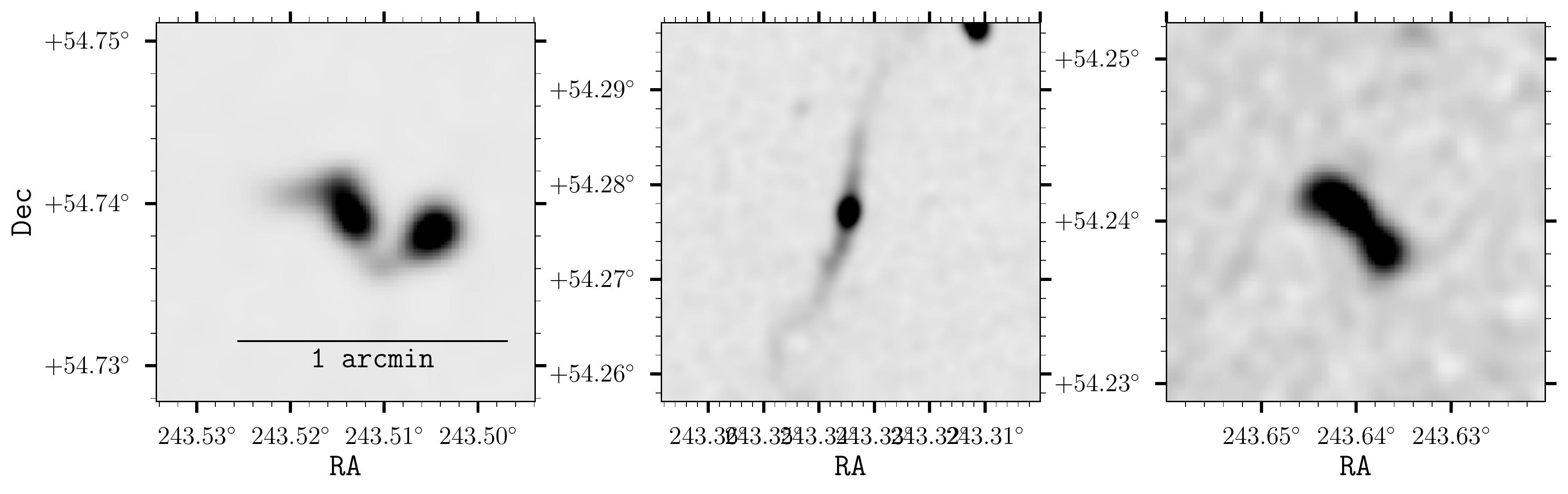}}
 \centerline{\includegraphics[width = 0.56\textwidth]{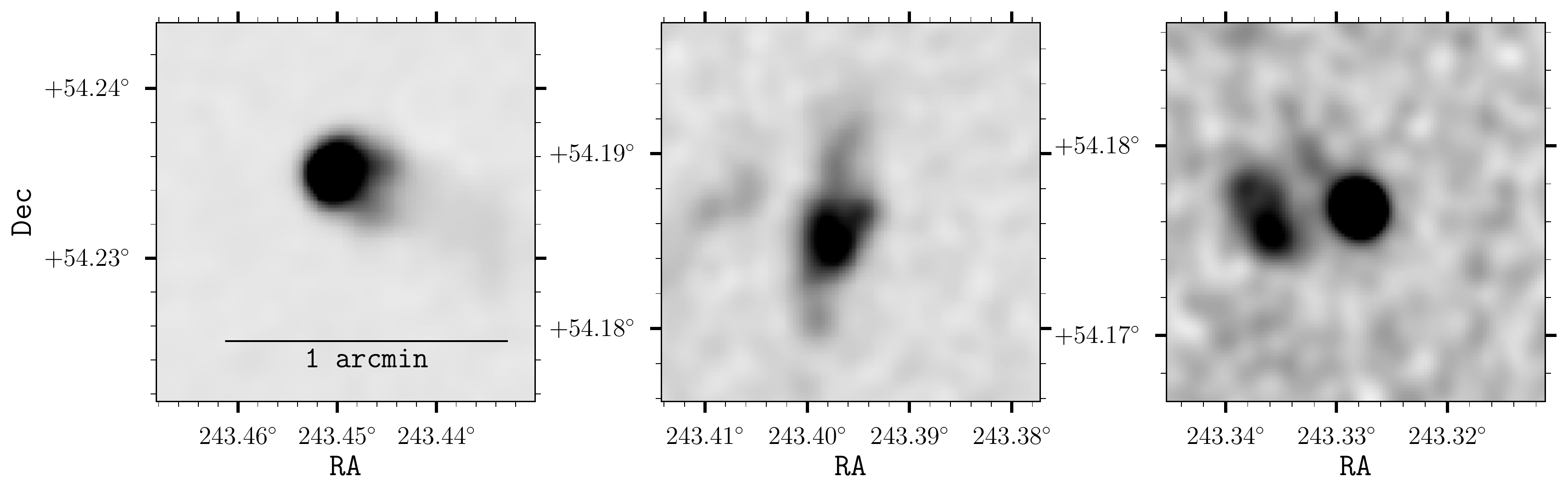}}
\caption{Postage stamps from the GMRT 610 MHz continuum mosaic image showing some extended radio sources.}
\label{source_size.fig} 
 \end{figure}

\begin{figure}
\centering
  \begin{tabular}{@{}cccc@{}}
    \includegraphics[width=.38\textwidth]{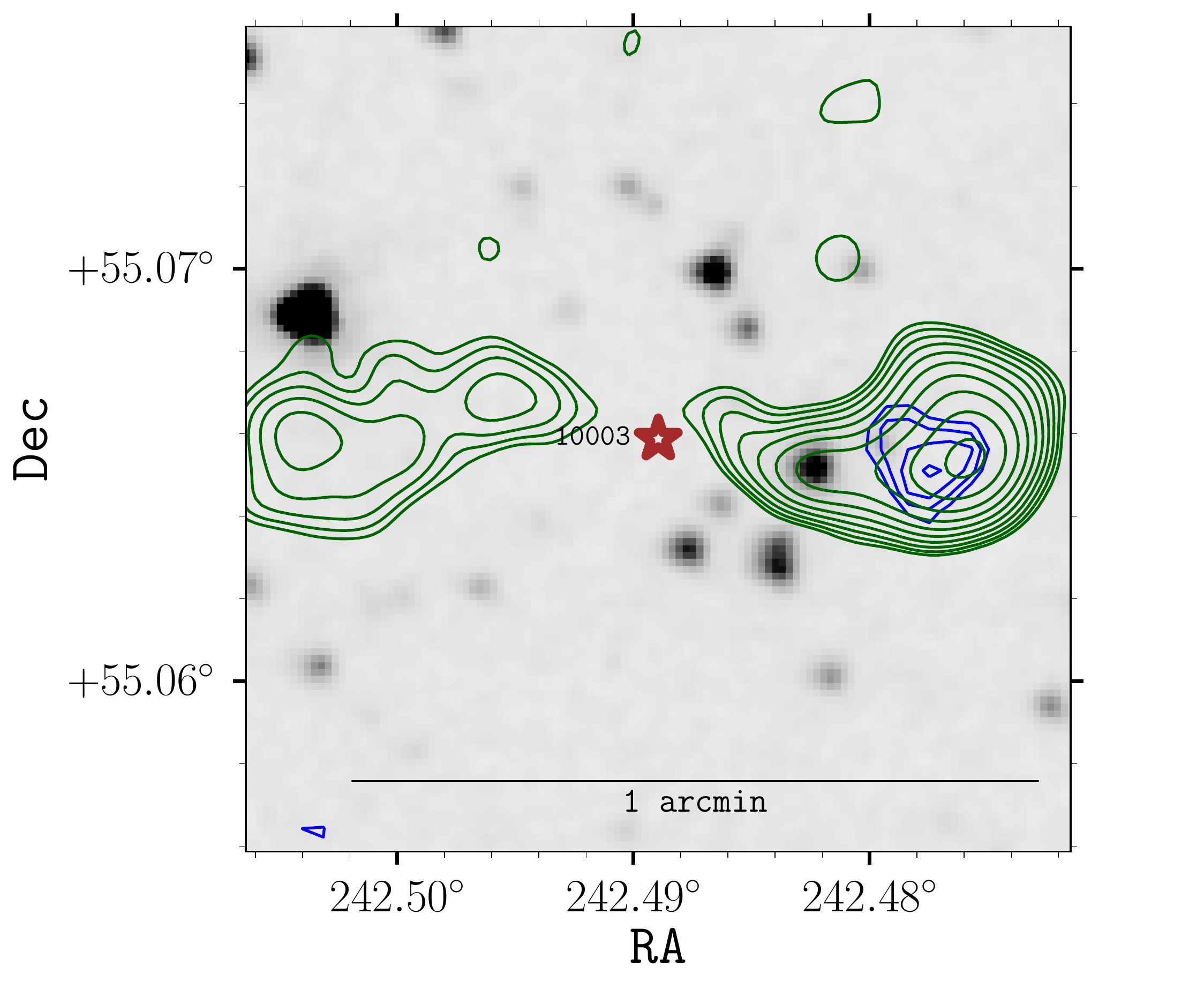} \\
    \includegraphics[width=.38\textwidth]{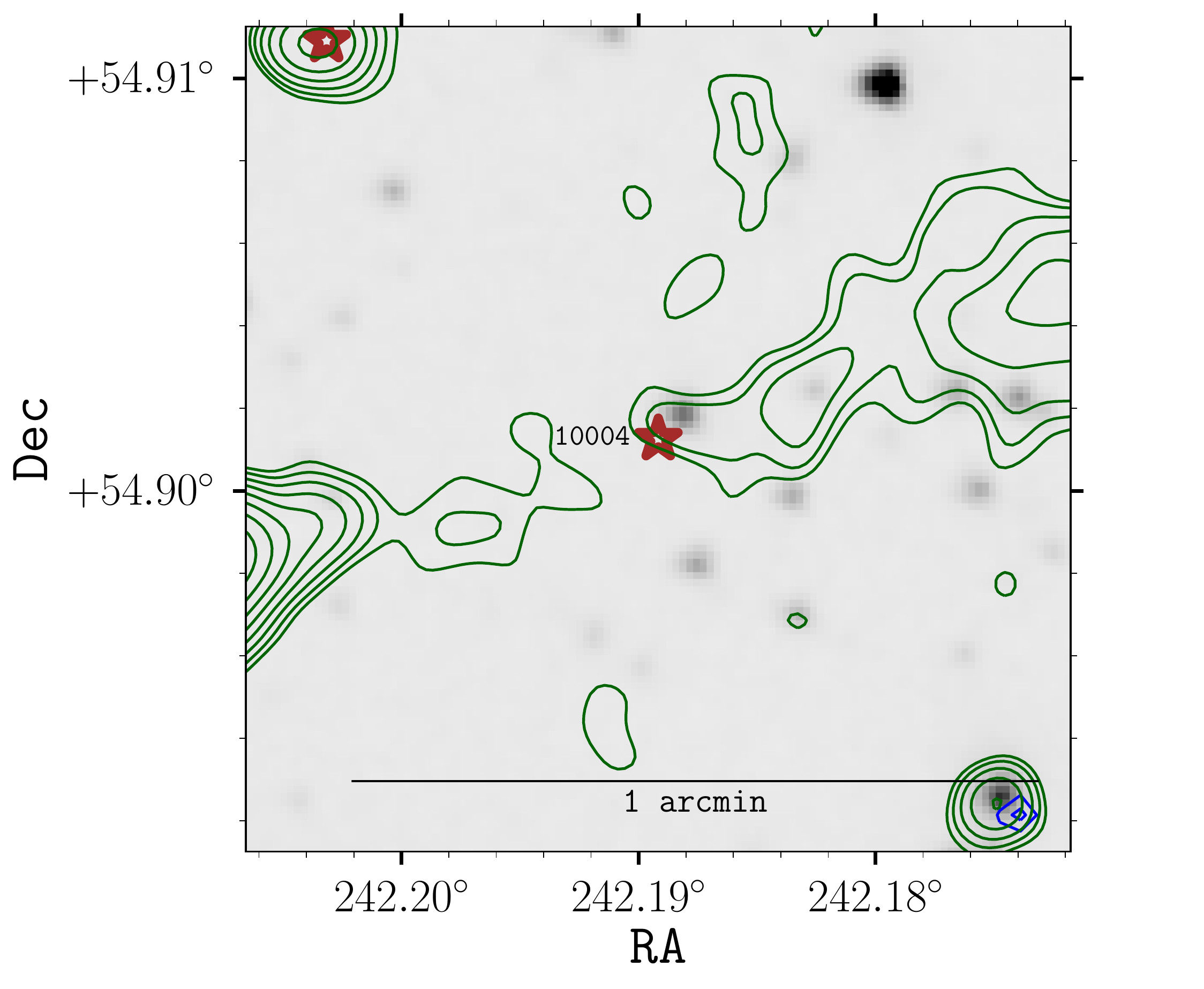} \\
     \includegraphics[width=.38\textwidth]{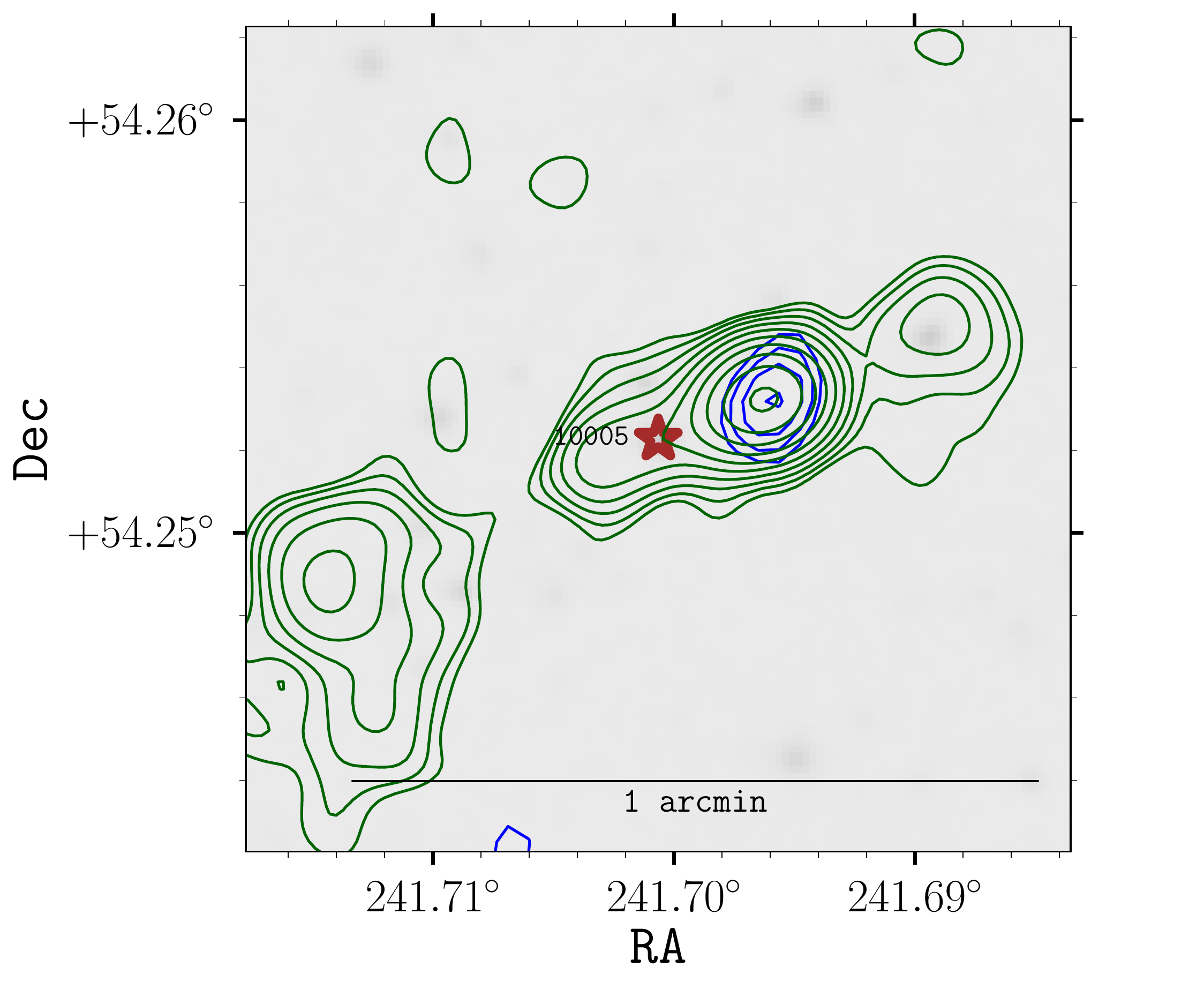} 
  \end{tabular}
  \caption{Postage stamps showing examples of multiple component sources in the GMRT 610 MHz catalogue. The greyscale shows IRAC band 1 and IRAC band 2 images respectively. The red stars show the central position of the GMRT source. The green contours represents the GMRT 610 MHz whereas the blue contours represents VLA FIRST. The contours levels are 1, 2, 3, 4, 5, and 6 $\rm{\sigma}$.}
\label{multi_servs.fig}   
\end{figure}

Figure~\ref{source_size.fig} shows examples of such sources (typically radio galaxies  or resolved star-forming disks). 
For the identification of these objects we make use of the Spitzer Extragalactic Representative Volume Survey (SERVS, \citealt{Mauduit2012}) which imaged $\rm{18 \ deg^{2}}$ using the IRAC1 3.6 $\rm{\mu m}$ and IRAC2 4.5 $\rm{\mu m}$ bands. SERVS overlaps with several other surveys from the optical, near- through far-infrared, sub-millimeter and radio. We overlaid radio contours on IRAC1 3.6 $\rm{\mu m}$ postage stamps. If we clearly identify a SERVS counterpart, we take the SERVS position to be the position of the radio source. Otherwise the radio source positions were determined by averaging the peak positions of the radio emission. The total and peak flux densities were estimated by summing the flux density inside regions guided by contours. 
Figure~\ref{multi_servs.fig} shows examples of SERVS+radio cutouts of extended or otherwise complex sources in the catalogue (i.e. more examples of these sources are shown in Figure~\ref{fig:postage_stamps}). 

Following the above process we produced a curated catalogue of 4290 sources which we used for our science analyses.



\section{SOURCE COUNTS}\label{sourcecounts.sec}
We derived number counts in the  EN1 using the curated catalogue shown in Table~\ref{catalogue.tab}. The radio number counts require no additional data but nevertheless provide very useful information, as their shape is tightly related to the evolutionary properties of the sources and also to the geometry of the Universe \cite{Padovani2016}.
The differential number counts, $\rm{dN/dS}$, were calculated using the observed number of sources per bin of flux density, $N$, corrected for the estimated number of false detections $N_f$, divided by the bin width ($\Delta$S in Jy) and multiplied  by the weight, \textsc{w} (which incorporates the efficiency, resolution bias $\rm{C_{R}}$ and Eddington bias $\rm{C_{Edd}}$)

\begin{equation}\label{eq:dnds}
\rm{\frac{dN}{dS}\,=\,\frac{N - N_f }{\Delta S}\,\times \textsc{w}}
\end{equation}

In this section, we discuss how we derive our source counts alongside our treatment of efficiency, resolution bias and Eddington bias. 

\subsection{Source sizes}\label{src_sizes.sec}
\begin{figure}
 \centering
 \centerline{\includegraphics[width = 0.5\textwidth]{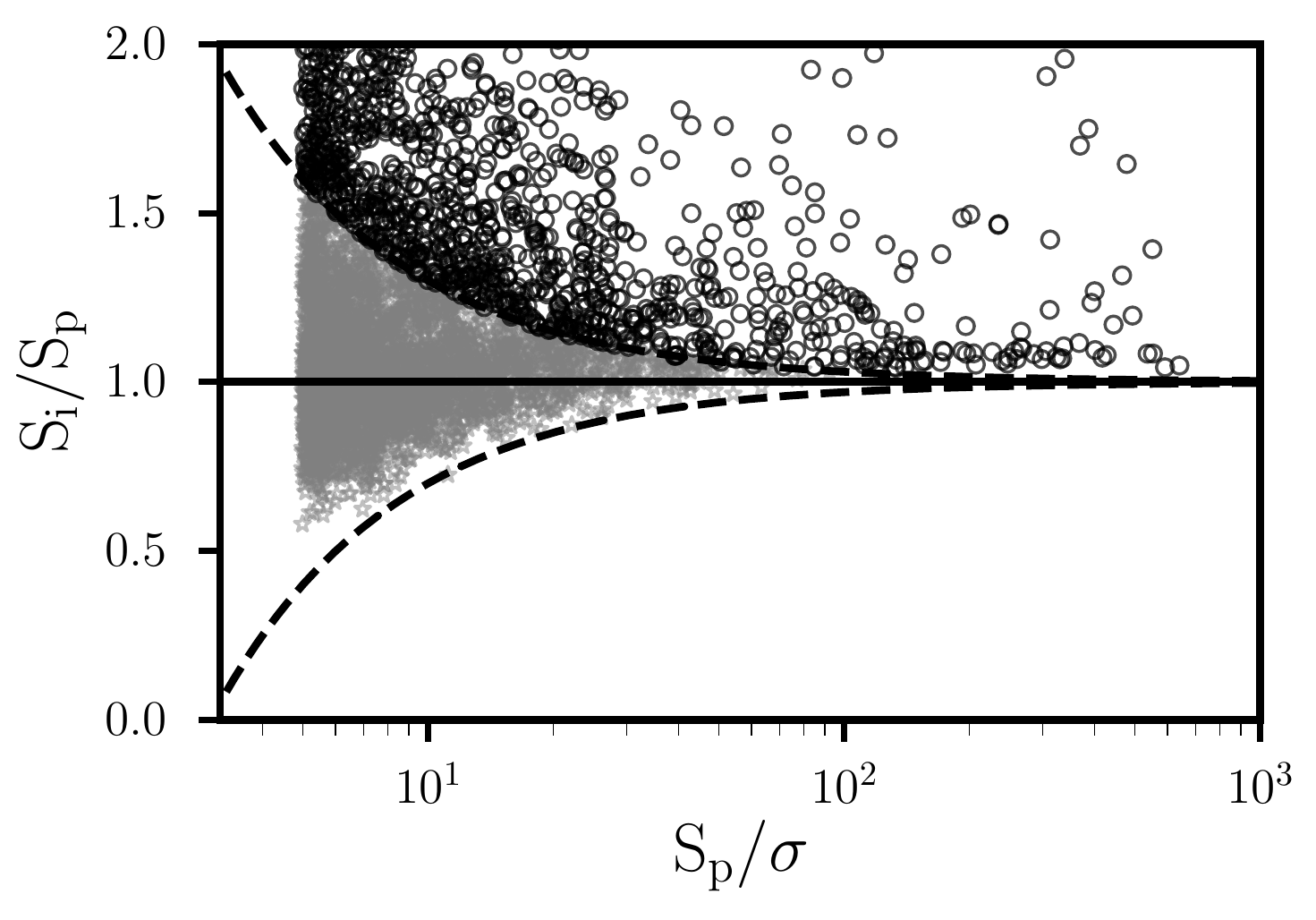}}
\caption{Ratio of the integrated flux density to peak flux density as a function of signal-to-noise ratio ($\mathrm{S_{p}/\sigma}$). Sources which are classified as
unresolved (grey stars) and resolved (open black circles) during the source-fitting procedure. {\bf The solid line} is at $\mathrm{S_{i}/ S_{p}\,=\,1}$.}
\label{fig:source_size} 
 \end{figure}

The flux density ratio may be used to discriminate between
point-like and extended sources
(see \citealt{2001A&A...365..392P,2006A&A...457..517P}).
The ratio of the integrated to peak flux densities is shown as a function of signal-to-noise ratio in Figure~\ref{fig:source_size}, with sources classified as point-like and extended shown separately. To select the resolved components, we determined the lower envelope of the points in Figure~\ref{fig:source_size}, by fitting a functional form  that can be characterized by Equation~\ref{src_size}. 
Almost all of the points with $\rm{S_{i}/S_{p}\,<\,1}$ lie above the curve. Reflecting this curve above the $\rm{S_{i}/S_{p}\,=\,1}$ line (upper envelope in Figure~\ref{fig:source_size}) gives a list of all the sources which lie above the upper envelope and can be considered to be resolved. This analysis shows that about 29\% of the sources (1260/4290) are considered to be resolved.

\begin{equation}\label{src_size}
\centering
\rm{\frac{S_i}{S_p}\,=\,1.0\,\pm\, \frac{3}{SNR}}
\end{equation}
where $\rm{SNR\,=\,\frac{S_p}{rms}}$.
Sources below this locus are considered to be unresolved.
These resolved and unresolved components were flagged in the catalogue.
We use the peak flux density as recovered by \textsc{PyBDSF} in place of the integrated flux density for unresolved sources when deriving the differential source counts.

\begin{figure}
 \centering
 \centerline{\includegraphics[width = 0.65\textwidth]{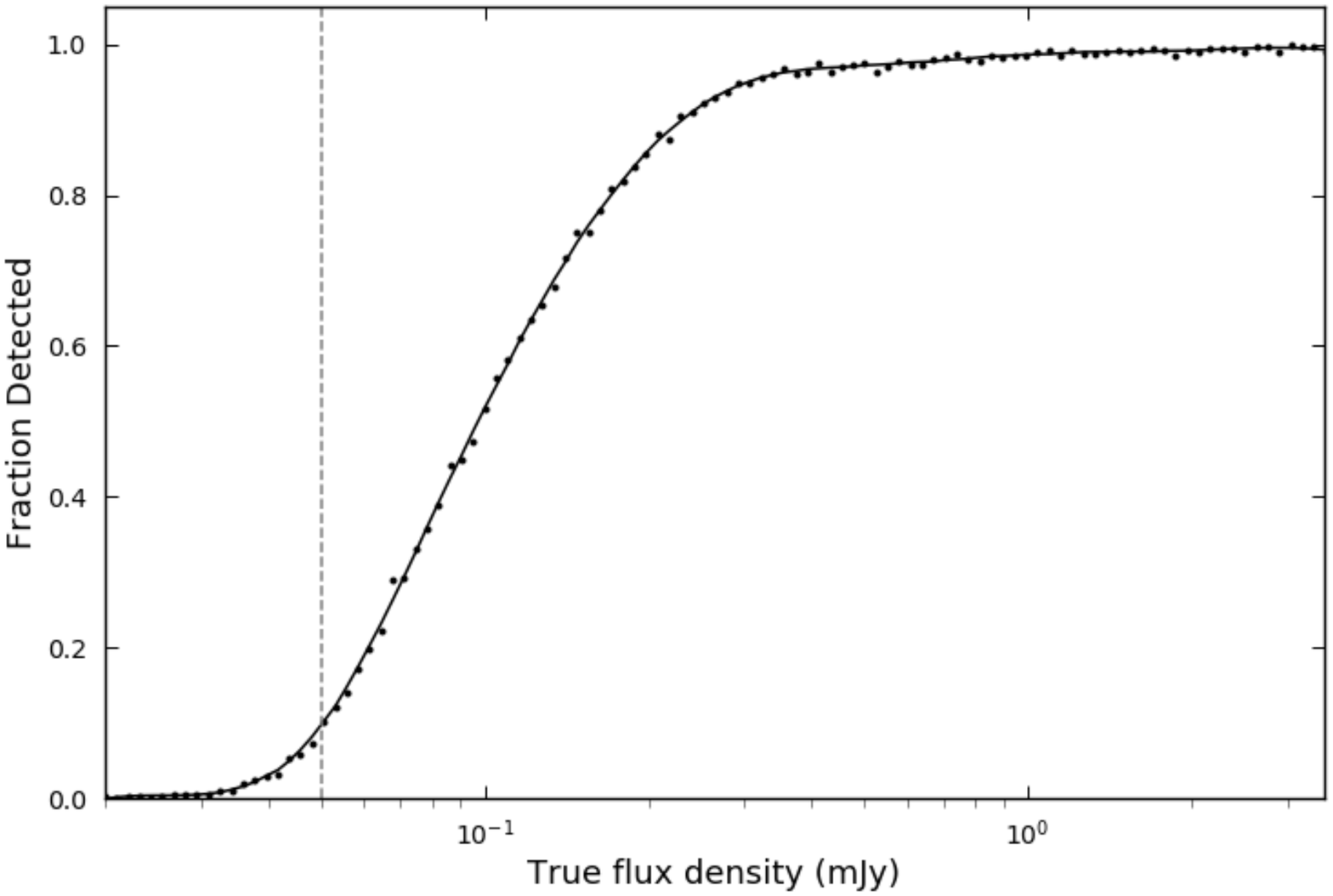}}
\caption{The fraction of simulated sources detected as a function of flux density illustrated by the blue solid curve. The solid line is a piece-wise 
polynomial spline interpolation of the data.
The vertical dashed line shows the approximate $5\sigma$ detection level of 50\,$\mu$Jy/beam.}
\label{completeness} 
 \end{figure}
 
\subsection{Reliability}
The reliability of a source catalogue is defined as the probability that all detected sources in the survey area above a certain  brightness detect limit are real sources and are not detections of artefacts or noise peaks (\citealt{2016MNRAS.460.2385W, doi:10.1093/mnras/stw2638}). 

We investigate the effect these may have on the false detection rate by running the \textsc{PyBDSF} algorithm with the same parameters we use to compile the source catalogue on an inverted image in exactly the same way as described in Section~\ref{Source finding}.
The source finding algorithm only detects positive peaks, therefore by inverting the image and running \textsc{PyBDSF} on the inverted map, any detections result from noise on the map \citet{doi:10.1093/mnras/stw2638}.
We detected 192 sources in the inverted image (compared to 6605 in the real image), giving a false detection rate of 2.5 per cent, which indicates, that the noise in the image is not entirely Gaussian. 
We corrected for false sources the source counts by subtracting the negative sources in each flux bin before calculating the counts. 


\subsection{Resolution bias}
The underestimation of source counts in a given flux density bin due to a resolved component having a lower peak flux density than an unresolved component with equivalent integrated flux density is defined as resolution bias (e.g. see \citealt{1987IAUS..124..545K,2001A&A...365..392P,2016MNRAS.460.2385W}). We calculate the approximate maximum size $\rm{\theta_{max}}$ a source could have for a given integrated flux density before it drops below the peak flux detection threshold. 
Using the relation below:

\begin{equation}\label{eq:res1}
\rm{\frac{S_{int}}{S_{peak}}\, = \, \frac{\theta_{maj}\theta_{min}}{b_{maj}b_{min}}}
\end{equation}
where $\rm{b_{min}}$ and $\rm{b_{maj}}$ are the synthesized beam axes, and $\rm{\theta_{min}}$ and $\rm{\theta_{maj}}$
are the deconvolved source axes, we estimate the maximum size a source of a given integrated flux density can have before falling below the peak flux detection threshold.

\begin{equation}\label{eq:res2}
\rm{\theta_{max} = \Bigg[(b_{maj}b_{min}) \frac{S_{int}}{5\sigma}\Bigg]^{0.5} }
\end{equation}
Combining the upper envelope for resolved sources defined in equation~\ref{src_size} (see Section~\ref{src_sizes.sec})  with equation~\ref{eq:res1} gives:
\begin{equation}\label{eq:res3}
\rm{\theta_{min} = \Bigg[(b_{maj}b_{min}) \Bigg(1+\frac{3}{SNR}\Bigg)\Bigg]^{0.5} }
\end{equation}
where $\rm{\theta_{min}}$ is the minimum angular size a source can have before it can be considered to be resolved as a function of its SNR (see \citealt{2016MNRAS.460.4433H}). We estimate the fraction of sources with deconvolved angular sizes larger than this $\rm{\theta_{max}}$ limit using the assumed true angular size distribution proposed by \cite{1990ASPC...10..389W}.

\begin{equation}\label{eq:res4}
\rm{h(>\theta) = \exp\Bigg[-\ln 2\Bigg(\frac{\theta_{lim}}{\theta_{med}}\Bigg)^{0.62}\Bigg] }
\end{equation}
where  $\rm{\theta_{lim}\, = \,max(\theta_{min},\theta_{max})}$
and $\rm{\theta_{med}\,=\,2\,S_{1.4GHz}^{0.3}}$ ( S is the flux at 1.4 GHz density in mJy, we have scaled the 1.4 GHz flux densities to 610 MHz with a spectral index of -0.8).

\begin{figure}
 \centering
 \centerline{\includegraphics[width = 0.50\textwidth]{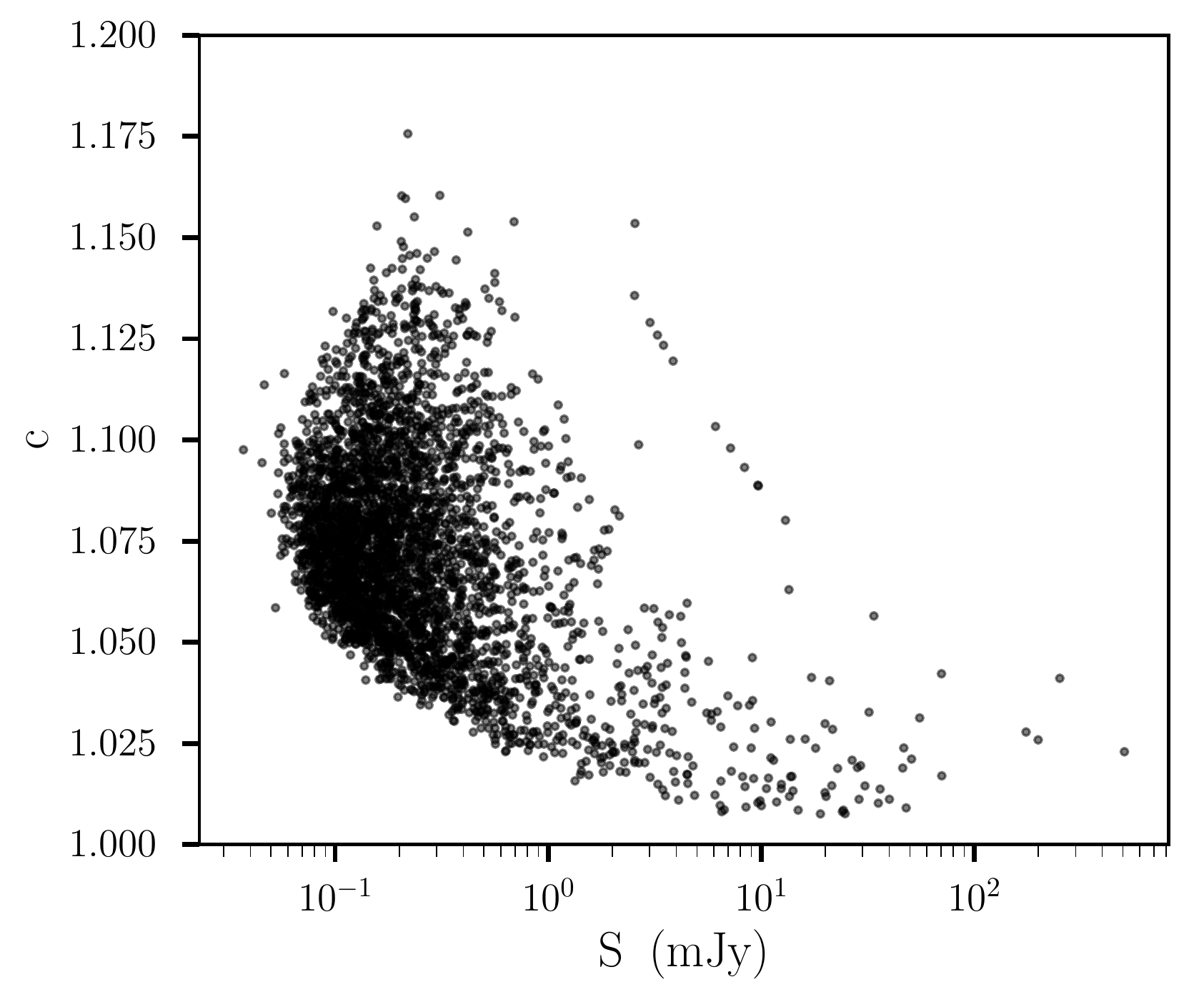}}
\caption{The resolution bias correction factor $\rm{\textit{c}_{R}\,=\,1/[1\,-\,h(>\,max)]}$ as a function of flux density.}
\label{fig:res_bias} 
 \end{figure}

 
The resolution bias correction factor $\rm{\textit{c}_{R}}$ for the counts is then given by:
\begin{equation}\label{eq:res5}
\rm{\textit{C}_{R}\,=\,\frac{1}{1\,-\,h(>\theta_{lim})}}
\end{equation} 
The correction factors calculated using the median size distributions is plotted as a function of flux density in Figure~\ref{fig:res_bias}. 

\subsection{Completeness and Eddington bias}

\cite{1913MNRAS..73..359E} showed that there was a significant bias in the measured number counts of stars even when the errors on the flux densities of the stars have the usual Gaussian distribution. This causes the apparent steepening of the observed source count by the intensity-dependent over-estimation of intensities, due to either system noise or confusion noise or both \citep{2015aska.confE.172Z}. This effect is more significant near the detection limit of a survey and could cause the number of observed sources to be slightly too high in the fainter bins.  To quantify the effect that Eddington bias has on source counts, previous work at higher frequency have semi-empirical methods. For example, \cite{2007MNRAS.378..995M}  used the best-fitting population model of the source count extrapolated to 60 $\mu$Jy as a prediction of the source counts below the detection limit. They subsequently derive counts from this population and use the difference between the recovered population and input model to quantify the Eddington bias.

To correct for both Eddington bias and the detection efficiency as a function of flux density we followed the approach outlined in \citet{2019MNRAS_Ishwara-Chandra}, which uses simulations to take into account the variation of the noise background of the mosaic image. The observed differential source counts can be related to the true source counts as 
\begin{equation}
\rm{\frac{dN_o(s')}{ds'}= \int_0^\infty \frac{dN_t(s)}{ds} \, \epsilon(s) \, p(s,s')\,ds}
\label{eqn:counts}
\end{equation}
Here $dN_o(s')/ds'$ is the observed count at observed flux densities $s'$, and $dN_t(s)/ds$ is the true source count at the true flux $s$. 
The function $p(s,s')$ is the normalized probability density function that a source at observed flux $s'$ is due to a source with true flux density $s$, and $\epsilon(s)$ is the probability that a source with true flux density,  $s$, will result in a detection - the completeness of the source catalogue versus true flux density.
We measured both function by inserting 3000 artificial point sources at a given true flux density at random positions into the residual map with the original sources removed. These sources populate the image with the same background noise and rms properties as the original source finding. The image was then searched for sources using the same parameters as for the real source list.
Figure~\ref{completeness} shows the result for $\epsilon(s)$. The field of view effect dominates the curve in Figure~\ref{completeness} since the analysis is incorporating the varying sensitivity limit across the field of view due to the GMRT primary beam. The effect of Eddington bias is clearly seen in the fact that sources with true flux well below the detection threshold have significant probability to produce detections.  
The combined completeness and Eddington bias correction is derived by iteratively inverting Equation~\ref{eqn:counts} to derive the correction factor that relates the true count to the observed count (see \citealt{2019MNRAS_Ishwara-Chandra} for details).


\subsection{The 610 MHz source counts}
\begin{figure*}
 \centering
 \centerline{\includegraphics[width = 0.9\textwidth]{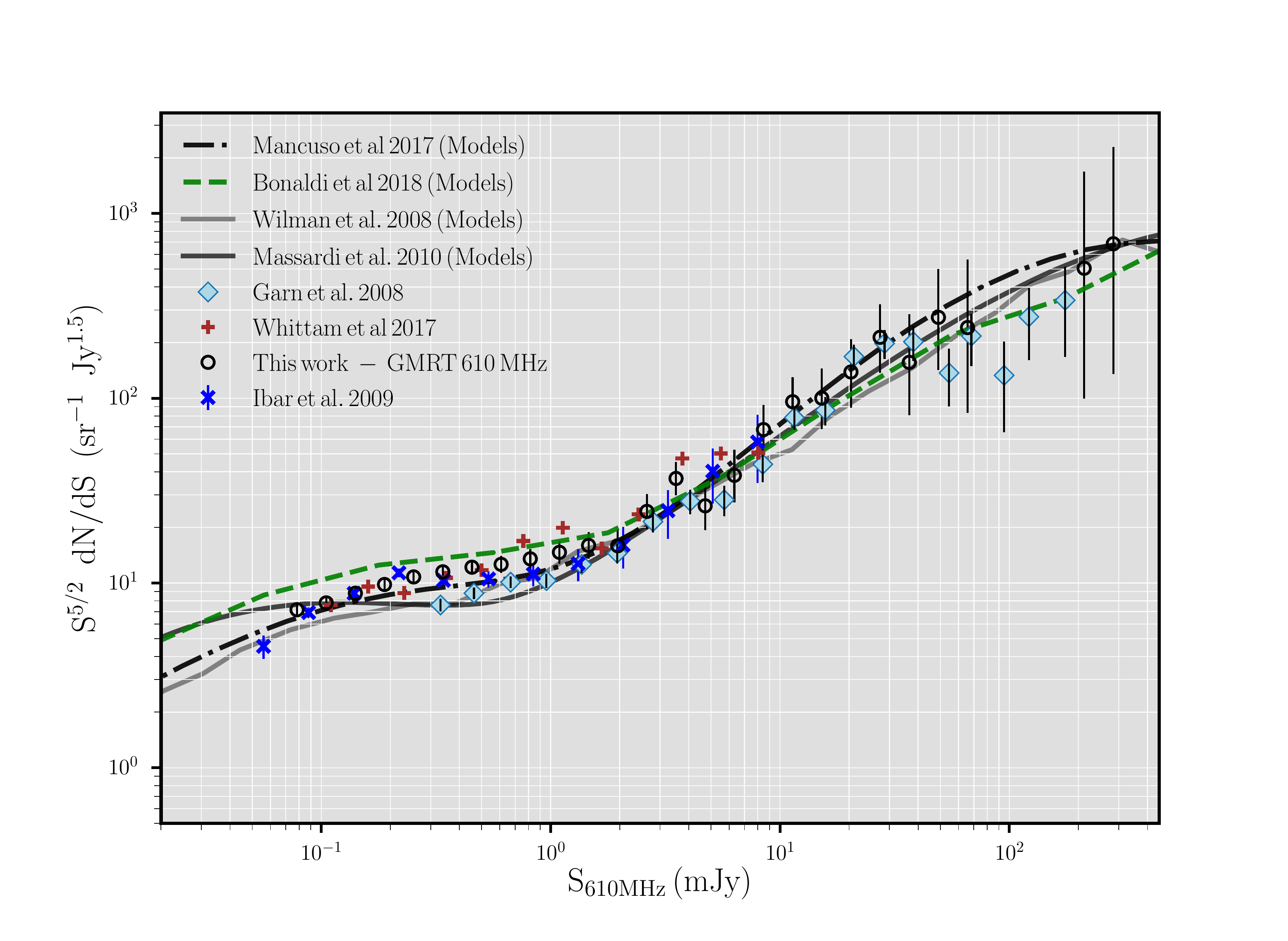}}
 \caption{Normalized 610-MHz differential source counts as derived from the catalogue discussed in this work (black points). Vertical bars represent Poissonian errors on the normalized counts. We compare with results from previous observations at 610 MHz including: \citet{2008MNRAS38375G} (filled blue diamonds), \citet{2009MNRAS.397..281I} (blue stars), \citet{doi:10.1093/mnras/stw2638} (brown pluses). We also compare against various models at 610 MHz including: \citet{2008MNRAS.388.1335W} (solid grey curve), \citet{2010MNRAS.404..532M} (solid black curve),  \citet{Mancuso2017} (dotted dashed black curve) and  \citet{Bonaldi2019} (dashed green curve).}
 \label{fig:counts} 
\end{figure*}

To compute the 610 MHz source counts, we used the integrated flux density if a source is classified as extended using the criteria described in Section~\ref{src_sizes.sec}. 
If a source is point-like, we instead use the peak flux density since this provides a better measure of the flux density of unresolved sources \citep{2009MNRAS.395..269S,doi:10.1093/mnras/stw2638}
We compute the weight which take into account the efficiency $\epsilon$(s), resolution bias,$\rm{C_{R}}$, and Eddington bias $\rm{C_{Edd}}$ given by the equation below:
$$
\rm{\textsc{w}\,=\,\frac{1}{\epsilon(s)}\times\,C_{R}\times\,C_{Edd}}
$$
Figure~\ref{fig:counts} illustrates the Euclidean-normalized differential source counts as derived from the catalogue discussed in this work (filled black points). The source counts
are tabulated in Table~\ref{src_cnt.tab}.
Uncertainties on the final normalized source counts are
propagated from the errors on the reliability and resolution bias correction factors and the Poisson
errors using the prescription of \citet{1986ApJ...303..336G} on the raw counts per bin. We do not add uncertainties associated with the Eddington bias correction due to the computational expense of running the full required simulation. The bin sizes are in {\bf linear} space and statistically independent. Each bin's upper limit is 1.34 times the lower limit. We note that for bins 75.264 - 100.854, 100.854-135.145 and 135.145-181.094 there are no sources in these bins hence the count is zero.

Many studies have observed a flattening in the source counts below 1 mJy at 1.4 GHz (\citealt{1984PhDT........89W, 1984Sci...225...23F,1984ApJ...287..461C}). This flattening has been later observed also at other frequencies, including 610 MHz \citep{2008MNRAS38375G,2010MNRAS.401L..53I,doi:10.1093/mnras/stw2638}. This present work confirms this flattening $\sim$ 1 mJy down to at least 100 $\mu$Jy (see Figure~\ref{fig:counts}). Below 100 $\mu$Jy 
our counts are probably less reliable, and we cannot entirely believe in the re-steepening we see. Our counts seems to be in better agreement with the later, but we caveat that existing radio source count models are better constrained at higher frequency (1.4 GHz) and their extrapolation to much lower frequency heavily relies on the assumptions on the spectral index source distribution. 

In this work we extend the number counts down to very faint 610 MHz flux densities whilst maintaining good agreement with simulations of previous studies at this frequency. Simulated counts by \cite{2010MNRAS.404..532M} are also shown in solid black curve in Figure~\ref{fig:counts}. \cite{Mancuso2017} investigated the astrophysics of radio-emitting star-forming galaxies and active galactic nuclei (AGNs) and explained their statistical properties in the radio
band including number counts. The dotted dashed black curve in Figure~\ref{fig:counts} show the \cite{Mancuso2017} models we compare to our work. The dashed green curve in Figure~\ref{fig:counts} shows the simulated 610 MHz counts from the Tiered Radio Extragalactic Continuum Simulation (T-RECS) by \cite{Bonaldi2019}. This new simulation of the radio sky in continuum models two main populations of radio galaxies: Active Galactic Nuclei (AGNs) and Star-Forming Galaxies (SFGs), and corresponding sub-populations, over the 150 MHz - 20 GHz range.

\begin{table*}
 \centering
 \caption{610 MHz radio source counts within the EN1 $\rm{1.864\,deg^{2}}$ field, normalized to Euclidean geometry. We chose fixed \textbf{(in linear space)} bin sizes and non-overlapping (statistically independent) bins.}
 \begin{threeparttable}
 \begin{tabular}{|c|c|c|c|c|c|c|}
 \hline
 \hline
$\rm{S_{range}}$& $\rm{S_{width}}$ &  $\rm{S_{mid}}$& Area &$\rm{\mathcal{C}_{Edd}}$ &N & Count  \\
($\rm{mJy}$)&($\rm{mJy}$)&($\rm{mJy}$)&($\rm{deg^{2}}$)&&&($\rm{Jy^{1.5}\,sr^{-1}}$)\\
(1)&(2)&(3)&(4)&(5)&(6)&(7)
\\
 \hline
0.067 - 0.090 &0.023  &0.078&0.667&3.041 &559 &$7.165^{+0.552}_{-0.538}$\\

0.090 - 0.120 &0.030  &0.105&1.009&1.981 &612 &$7.802^{+0.583}_{-0.571}$\\

0.120 - 0.161 &0.041  &0.141&1.313&1.326& 670 &$8.802^{+0.585}_{-0.572}$\\

0.161 - 0.216 &0.055  &0.189&1.546&1.148 &566&$9.814^{+0.613}_{-0.596}$\\

0.216 - 0.289 &0.073 &0.253&1.705&1.032 & 447&$10.787^{+0.713}_{-0.686}$\\

0.289 - 0.388 &0.098  &0.339&1.783&1.016&316 &$11.490^{+0.832}_{-0.796}$\\

0.388 - 0.520 &0.132  &0.454&1.807&0.926&234 &$12.173^{+0.961}_{-0.909}$\\

0.520 - 0.696 &0.177 &0.608&1.818&0.987 &150&$12.612^{+1.234}_{-1.141}$\\

0.696 - 0.933 &0.237 &0.815&1.831&1.105 &90&$13.511^{+1.576}_{-1.426}$\\

0.933 - 1.251 &0.317 &1.092&1.840&1.000 &68&$14.653^{+2.004}_{-1.767}$\\

1.251 - 1.676 &0.425 &1.463&1.846&1.000 &49&$15.969^{+2.607}_{-2.281}$\\

1.676 - 2.246 &0.570 &1.961&1.848&1.000 &32&$15.906^{+3.330}_{-2.784}$\\

2.246 - 3.009 &0.764 &2.627&1.854&1.000 &31&$24.397^{+5.448}_{-4.582}$\\

3.009 - 4.032 &1.023 &3.521&1.859&1.000 &31&$36.792^{+7.833}_{-6.539}$\\

4.032 - 5.403 &1.371  &4.718&1.864&1.000 & 16&$26.159^{+8.338}_{-6.539}$\\

5.403 - 7.240 &1.837  &6.322&1.864&1.000&13&$38.250^{+13.829}_{-10.592}$\\

7.240 - 9.702 &2.462  &8.471&1.864&1.000&14&$67.653^{+23.195}_{-17.879}$\\

9.702 - 13.001 &3.299  &11.351&1.864&1.000&14&$95.745^{+32.827}_{-25.304}$\\
13.001 - 17.421 &4.420  &15.211&1.864&1.000&10&$100.053^{+43.022}_{-31.017}$\\

17.421 - 23.344 &5.923  &20.382&1.864&1.000 &8&$138.663^{+67.598}_{-48.532}$\\

23.344 - 31.281 &7.937 &27.312 &1.864&1.000&8&$213.640^{+104.149}_{-74.774}$\\

31.281 - 41.916 &10.635 &36.598&1.864&1.000&4&$156.313^{+125.050}_{-74.249}$\\

41.916 - 56.168 &14.251 &49.042&1.864&1.000 &4&$274.038^{+219.231}_{-130.168}$\\

56.168 - 75.264&19.097&65.716&1.864&1.000 &2&$241.168^{+313.518}_{-156.759}$\\

75.264 - 100.854&25.590&88.059&-&-&-&-\\

100.854 - 135.145&34.290&118.000&-&-&-&-\\

135.145 - 181.094&45.949&158.119&-&-&-&-\\

181.094 - 242.666&61.572&211.880&1.864&1.000& 1&$505.120^{+1161.775}_{-404.096}$\\
242.666 - 325.173&82.506&283.919&1.864&1.000&1&$685.387^{+1576.389}_{-548.309}$\\
\hline
\end{tabular}
\begin{tablenotes}
\item [a] The listed counts were corrected for completeness and bias corrections (Resolution and Eddington Bias) (see text for details).
\item (1) the flux density bins.
\item (2) the width of the flux density bins.
\item (3) the central flux density of the bin.
\item (4) the effective area corresponding to the bin centre.
\item (5) the Eddington bias correction factor.
\item (6) the number of sources in each flux density bin.
\item (7) the corrected normalized source counts.

\end{tablenotes}
\end{threeparttable}
\label{src_cnt.tab} 
\end{table*} 
 
\section{Multi-Wavelength Cross-Identification}\label{crossmatch.sec}
\subsection{Cross-matching}\label{crossmatch.sec_1}
One advantage of the ELAIS N1 field is the wealth of multi-wavelength data publicly available in the field to study the properties of radio sources. Most of this public data has been homogeneized as part of the SERVS Data Fusion project\footnote{\url{http://www.mattiavaccari.net/df}} (\citealt{Vaccari2010,Vaccari2015}) and of the
The Herschel Extragalactic Legacy Project \citep{Vaccari2016}\footnote{\url{https://herschel.sussex.ac.uk}}.

To determine their multi-wavelength counterparts, we first matched GMRT radio sources against  the Spitzer Extragalactic Representative Volume Survey SERVS DR2 \citep{Mauduit2012,Vaccari2015} positions using a variable search radius equal to three times the combined astrometric error. Where a SERVS match was not found, we used the UKIRT Infrared Deep Sky Survey (UKIDSS) Deep Extragalactic Survey (DXS) DR10Plus data release \citep{2007MNRAS.379.1599L}. Both the SERVS and UKIDSS catalogues were astrometrically calibrated against 2MASS, which provide a dense and accurate astrometric reference frame. Radio positional errors for individual sources from \textsc{PyBDSF} source finder \citep{2015ascl.soft02007M} are typically a few a tenths of an arcsecond. We computed the median astrometric offsets between the GMRT and SERVS catalogues to be $\mathrm{+0.539 \pm 0.420}$ in RA and $\mathrm{-0.327 \pm 0.422}$ arcsec in RA and DEC respectively from an initial cross-matching. We applied these corrections to the radio positions before performing a second cross-matching. We measured a median astrometric offsets for the second cross-matching to be $\mathrm{+0.055 \pm 0.447}$ in RA and $\mathrm{-0.026 \pm 0.435}$ in DEC. This correction was then applied to the radio positions for a final cross-matching. The radio positions within our final catalogue
in Table~\ref{catalogue.tab} were corrected following this procedure are are thus ultimately also registered against 2MASS.

Virtually all cases where a match was found resulted in a unique identification, given the sub-arcsecond accuracy of the positions. For all GMRT sources with a match in SERVS/UKIDSS we determined multi-wavelength properties using the SERVS Data Fusion workflow, i.e. matching all ancillary catalogues with a search radius of 1 arcsec against the SERVS/UKIDSS position. This ancillary data include IRAC1234 and MIPS 24 $\mu$m photometry from SWIRE \citep{Lonsdale2003}, PACS and SPIRE photometry from HerMES \citep{Oliver2012} and redshift information.
Table~\ref{catalogue.tab} shows a sample of ten rows and a few selected columns from the curated catalogue with multi-wavelength properties, which is available electronically in its entirety.

\begin{table*}
  \centering
  \caption{Sample of the source catalogue of GMRT 610 MHz sources. The columns are described in the text.}
  \label{catalogue.tab}
   \begin{threeparttable}
  \begin{tabular}{|c|c|c|c|c|c|c|c|c|c|c|c|}
 \hline
 \hline
    $\rm{ID}$  & $\rm{R.A}$ & $\rm{\sigma_{R.A}}$ & Dec& $\rm{\sigma_{Dec}}$ & $\rm{S_{int}}$ & $\rm{\sigma_{S_{int}}}$ &$\rm{S_{peak}}$ & $\rm{\sigma_{S_{peak}}}$&RMS&$\rm{S_{code}}$&Type\\ & \rm{ [deg]}  & $\rm{[arcsec]}$ & [deg]&$\rm{[arcsec]}$ &  $\rm{[mJy]}$ &$\rm{[mJy]}$ &$\rm{[mJy]}$ & $\rm{ [mJy]}$&$\rm{[mJy]}$ & &\\ 
    (1) &(2)&(3)&(4)&(5)&(6)&(7)&(8)&(9)&(10)&(11)&(12) \\ \hline
    1  &243.801805  &0.35  &54.621228  &0.36   &0.2610  &0.0555  & 0.2253  &0.0290  &0.0282 &S&P\\
    2  &243.810522  &0.04  &54.993388  &0.04   &3.2388  &0.0817  & 2.9474  & 0.0443  &0.0435 & S&E \\
    3  &243.797445 &0.12  &54.588408  &0.10   &0.7786  &0.0478  & 0.6214   &0.0237  &0.0228  &S&E\\
    4  &243.798035  &0.40  &54.594328  &0.57   &0.1594  &0.0460   &0.1432   &0.0245  &0.0242  &S&P\\
    5  &243.802174  &0.46  &54.764238  &0.74   &0.2206  &0.0654   &0.1583   &0.0298  &0.0283  &S&P\\
    6  &243.787447  &0.18  &54.310108 &0.16   &0.9742  &0.0871   &0.7077   &0.0404  &0.0383  &S&E\\
    7  &243.801123  &0.36  &54.782338 &0.30   &0.1354  &0.0379   &0.1773   &0.0260  &0.0283  &S&P\\
    8  &243.793715  &0.34  &54.539598  &0.30   &0.1761  &0.0438   &0.2095   &0.0283  &0.0297  &S&P\\
    9  &243.808151 &0.43  &55.138628  &0.64   &0.7233  &0.1393   &0.3626   &0.0488  &0.0461  &S&E \\
    10  &243.795284  &0.46  &54.718948  &0.45   &0.2611  &0.0627   &0.1945   &0.0296  &0.0280 &S&P\\

    \hline 
  \end{tabular}
  \begin{tablenotes}
  \item The catalogue columns are:
\item (1) - GMRT 610 MHz Source ID.
\item(2) and (3) - flux-weighted right ascension (RA) and uncertainty.
\item(4) and (5) - flux-weighted declination (Dec.) and uncertainty.
\item(6-7) - integrated source flux density and uncertainty.
\item (8-9) - peak flux density and uncertainty.
\item (10) - the average background rms value of the island.
\item (11) - Code that defines the source structure. 
S - a single-Gaussian source that is the only source in the island. M - a multi-Gaussian source.
\item (12) - defines a source as extended (E) or point source (P).
\end{tablenotes}
\end{threeparttable}
\end{table*}

\begin{table}
\centering
\caption{GMRT cross-matching statistics.}
\begin{threeparttable}
\begin{tabular}{lrr}
\hline
\hline
Category    & Size & Fraction ($\%$) \\
\hline
GMRT      		& 4290    & 100$\%$\\
Matched        	& 3689    & 92$\%$ \\ 
SERVS         	& 3689    & 86$\%$ \\ 
UKIDSS        	& 3542    & 83$\%$ \\ 
SWIRE IRAC1234  & 1623    & 43$\%$ \\ 
MIPS 24 $\mu$m	& 2714    & 63$\%$ \\ 
X-RAY           &  149    &  3$\%$ \\
SPEC-Z$^{\rm a}$&  834    & 19$\%$ \\
PHOTZ-HSC$^{\rm b}$& 2885    & 67$\%$ \\
PHOTZ-SWIRE$^{\rm c}$&  907    & 21$\%$ \\ 
PHOTZ-HELP$^{\rm d}$& 1834    & 43$\%$ \\
REDSHIFT$^{\rm e}$  & 3105    & 72$\%$ \\
CLASS $^{\rm f}$ & 3490    & 81$\%$ \\
REDSHIFT \& CLASS$^{\rm g}$ & 2304  & 54$\%$ \\
\hline
\end{tabular}
\begin{tablenotes}
\item [a] spectroscopic redshifts
\item [b] Hyper Suprime-Cam (HSC) Photometric Redshift Catalogue \citep{Tanaka2018}.
\item [c]  SWIRE Revised Photometric Redshift Catalogue \citep{RowanRobinson2008,RowanRobinson2013}.
\item [d] HELP Photometric Redshift Catalogue \citep{Duncan2018}
\item [e]  The union of SPECZ, PHOTOZ-HSC, PHOTOZ-SWIRE and PHOTOZ-HELP.
\item [f] All sources with at least one multi-wavelength classification diagnostic.
\item [g] All sources with at least one multi-wavelength classification diagnostic and redshift association.
\end{tablenotes}
\end{threeparttable}
\label{tab_matches} 
\end{table}
 
Table~\ref{tab_matches} summarizes the multi-wavelength and redshift information available for the cross-matched GMRT sources, included in the SERVS Data Fusion catalogue. Accounting for multiple sources the final number of our GMRT 610 MHz sources is 4290. The redshift information is discussed in more detail in the following section.

\subsection{Redshifts}\label{redshift.sec}
We have combined the spectroscopic redshift compilations from the Spitzer Data Fusion\footnote{\url{http://mattiavaccari.net/df/specz}} and HELP\footnote{\url{http://hedam.lam.fr/HELP/dataproducts/dmu23/}} projects to collect spectroscopic redshifts. The majority of the spectroscopic redshifts for our sample were obtained with the Baryon Oscillation Spectroscopic Survey (BOSS) \citep{Eisenstein2011}. This is supplemented by a small number of redshifts available from the literature and from SWIRE/HerMES spectroscopic follow-up programmes. For sources where a spectroscopic redshift was not available, we use photometric redshift estimates from the Hyper Suprime-Cam (HSC) project \citep{Tanaka2018}, the SWIRE project \citep{RowanRobinson2008,RowanRobinson2013} and the HELP project \citep{Duncan2018}.

\begin{figure}
\centering
\includegraphics[width = 0.5\textwidth]{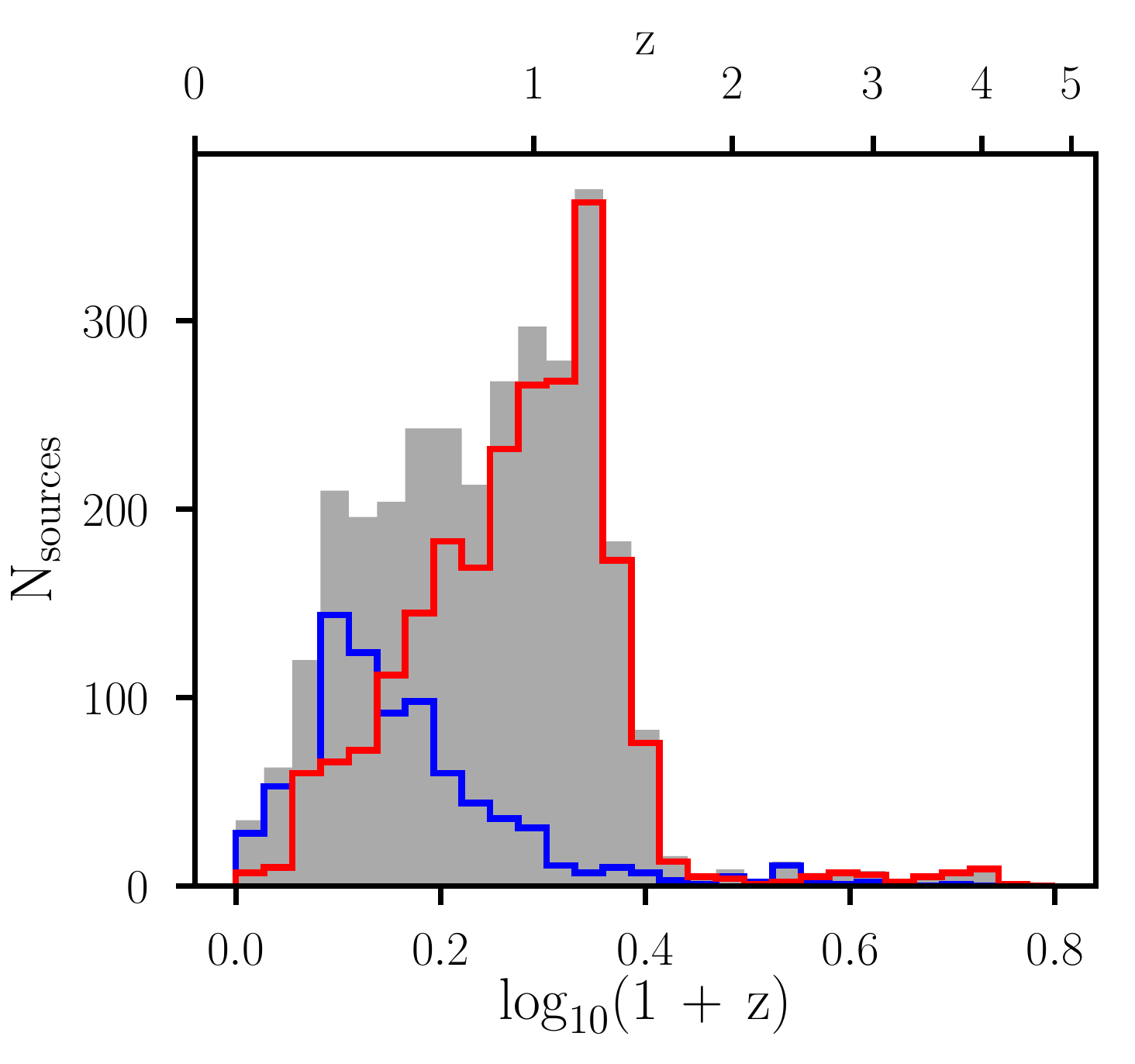}
\caption{Redshift distribution for the GMRT 610 MHz sources. The grey histogram represents all the redshifts (i.e. both spectroscopic and photometric redshifts). The blue and red histograms represents spectroscopic and photometric redshifts respectively.}
\label{fig:z_dist} 
\end{figure}

\begin{figure}
 \centering
 \centerline{\includegraphics[width = 0.55\textwidth]{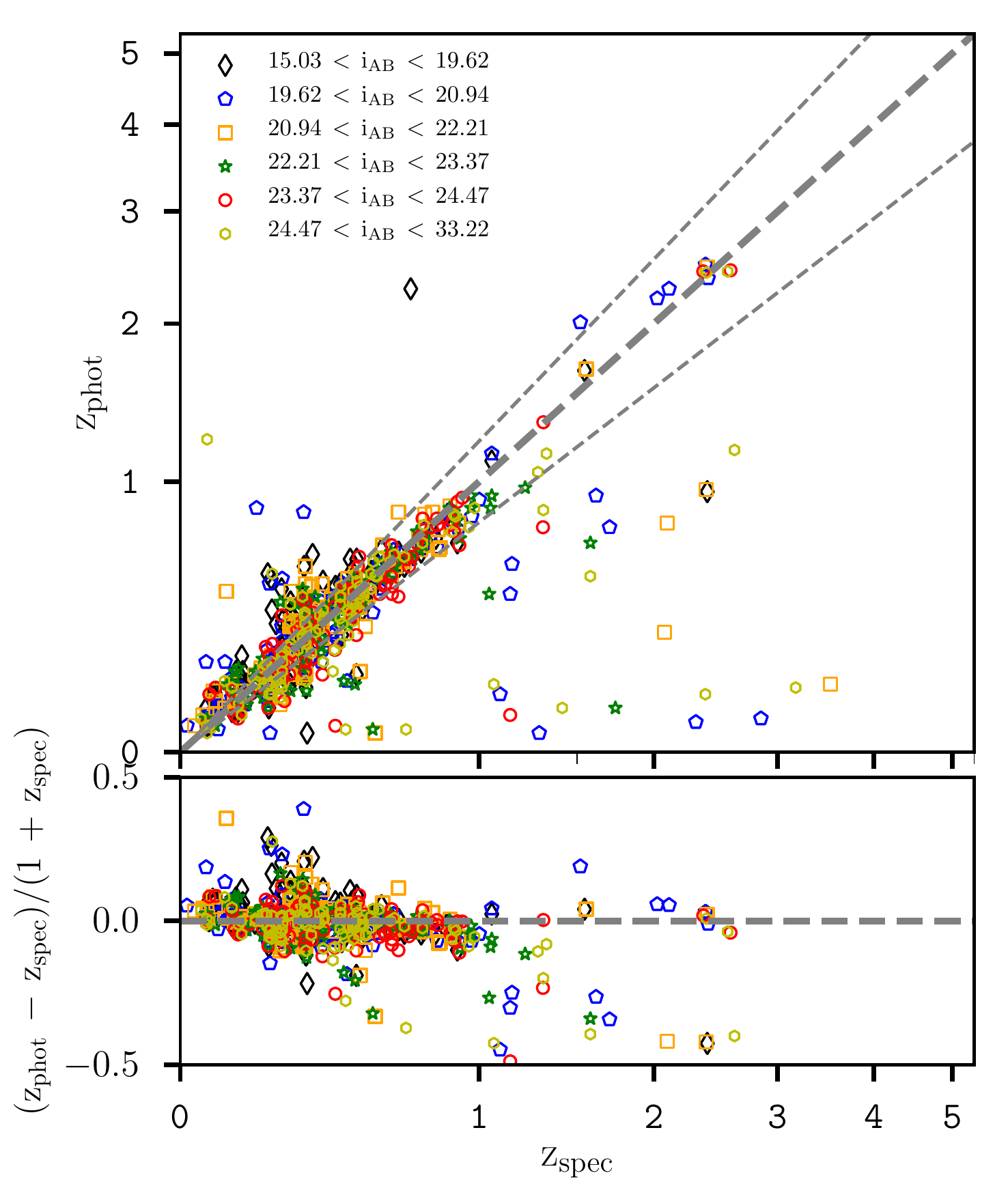}}
\caption{Comparison between photometric and spectroscopic redshifts as a function of $\rm{i_{AB}}$ magnitude in top panel. The dashed grey line corresponds to $\rm{ z_{spec} \ = \ z_{phot}}$. The double dashed  lines show $\rm{z_{phot}\,=\,z_{spec} \pm \,0.2(1+z_{spec}) }$. The lower panel show $\rm{(z_{phot}\,-\,z_{spec})/(1\,+\,z_{spec})}$ vs $\rm{z_{spec}}$ as a function of $\rm{i_{AB}}$ magnitude.}
\label{zphot_vs_zspec} 
 \end{figure}

The redshift distribution is shown in Figure~\ref{fig:z_dist} with the blue histogram representing spectroscopic redshifts and red histogram for photometric redshifts.
We estimate the precision of the photometric redshifts using the normalized median absolute (i.e. deviation $\rm{\sigma_{NMAD}}$) \citep{10.2307/3182748} given by $\rm{1.48\times median(|\Delta z|)/(1+z_{spec})}$. The second metric we estimate is the outlier fraction, $\mathrm{O}_{f}$, defined as $\rm{|\Delta z|/(1+z_{spec})\,>\,0.2}$ (see \citealt{2008ApJ...686.1503B,2013ApJ...775...93D,2016ApJS..224...24L,Duncan2018}). 
We use these metrics to explore the performance of the photometric redshift estimates relative to the measured spectroscopic redshift.
Figure~\ref{zphot_vs_zspec} (upper panel) compares the photometric and spectroscopic redshift for different bins of $\rm{i_{AB}}$ magnitude for the GMRT sample. The bottom panel of Figure~\ref{zphot_vs_zspec} presents {\bf $\rm{  (z_{phot}\,-\,z_{spec})/(1\,+\,z_{spec})}$ } as a  function  of  $\rm{z_{spec}}$. The  horizontal  line  in indicates where $\rm{  (z_{phot}\,-\,z_{spec})/(1\,+\,z_{spec})\,=\,0}$.  This  plot  clearly shows that the  fraction  of  outliers  increases  significantly  towards higher  values  of $\rm{z_{spec}}$ as a function of $\rm{i_{AB}}$ magnitude. The fraction of catastrophic 
failures varies from 3 to 9\% and the 
scatter, $\rm{NMAD}$ 
is nearly constant with an average value of $\rm{NMAD\,=\, 0.049}$ for the entire spectroscopic sample. The scatter does increase above $\rm{z_{spec}\, > \,1.5}$, where $\rm{\sigma_{NMAD}\,=\, 0.075}$.
Systematic deviations from the {\bf $\rm{z_{phot}\,=\,z_{spec}}$} line are very small at most redshifts, with the exception that $\rm{z_{phot}}$ underestimates $\rm{z_{spec}}$ at $\rm{z\,=\,1.0\, - \,1.4}$ by $\rm{\sim5\%}$.

\begin{table}
 \centering
 \caption{Photometric Redshift Performance for GMRT radio sources as a function of $i_{AB}$ optical magnitude.}
 \begin{tabular}{|c|c|c|}
 \hline
 \hline
$i_{AB}$ & $\rm{\sigma_{NMAD}}$ & $\mathrm{O}_{f}$ \\
 \hline
[15.03, 19.62]& 0.054    & 0.075\\

[19.62, 20.94]& 0.047    & 0.017\\
 
[20.94, 22.21]& 0.063    & 0.121\\

[22.21, 23.37]& 0.044    & 0.035\\

[23.37, 24.47]& 0.052    & 0.070\\

[24.47, 33.22]& 0.035    & 0.098\\
\hline
\end{tabular}
\label{redshift_quality} 
\end{table}

\subsection{AGN/SFG Diagnostics Overview}\label{diagnostics.sec}
Following \cite{2017MNRAS.468.1156O}, we have carried out a multi-wavelength study using optical, X-ray, infrared and radio diagnostics to search for evidence of AGN-driven activity in our sample. The total number of sources with redshifts for which we can define at least one AGN indicator is 2305 (i.e. $\sim\,54\%$ of the whole sample and $\sim\,74\%$ of the subsample with redshifts). The AGN diagnostics we employed are described below:

\begin{enumerate}
  \item[1.] Radio power: we classify sources as RL AGNs based on a radio luminosity cut of $L_{\rm 1.4\,GHz} \ge 10^{25}$  W\,Hz$^{-1}$ (e.g. \citealt{2007ApJ...667L..17S, 2007ApJ...656..680J, 2008ApJ...683..659S}). We converted the  610 MHz radio flux densities to rest-frame 1.4\,GHz effective luminosities assuming a radio spectral 
index of $\alpha = - 0.7$ (i.e. $\rm{S(\nu) \propto \nu^{\alpha}}$) \citep{2010MNRAS.401L..53I}. \footnote{$
L_{\rm 1.4\,GHz} \  = 4{\rm \pi} d^{2}_{\rm lum} {\frac{S_{\rm 1.4\,GHz}} {(1 + z) ^{1 +\alpha}}}\, ,\mathrm{where} \, S_{\rm 1.4\,GHz}\, = \,\left(\frac{1.4}{0.61}\right)^{\alpha}\,S_{\rm 0.6\,GHz} $}
 
 \item[2.] Mid-Infrared to radio flux ratio: following the \cite{2013MNRAS.436.3759B}, we compute $q_{24\mu \rm m}$ for the radio sources with MIPS 24$\rm{{\mu m}}$ detections and redshifts. This is then compared to the redshifted $q_{24\mu {\rm m}}$ value for the M82 local standard starburst galaxy template. If $q_{24\mu \rm m}$ is lower than the one expected for M82  (i.e. below $\rm{-2\sigma}$, $\rm{\sigma\,=\,0.35}$ average spread for local sources by \cite{2010ApJ...714L.190S}), the sources are considered as a RL AGN. 
 
 \item[3.]  X-ray luminosity: we classify a source as an AGN 
when $L_{\rm x} > 10^{42}$ erg\,s$^{-1}$  following e.g. \cite{2004ApJS..155..271S}. 
\footnote{
$L_{\rm x}=4 {\rm \pi} S_{\rm x} d_{\rm L}^{2} (1 +z)^{2 - \gamma} 
$ , where we fixed the the photon-index to the commonly observed value of $\rm{\gamma} = 1.8$ \citep{2008A&A...485..417D, 2014MNRAS.445.3557V}.}

 \item[4.] BOSS AGN spectroscopic classification: we use the BOSS CLASS and SUBCLASS parameters, as detailed by 
\cite{2012AJ....144..144B}, to classify the GMRT sources with BOSS identifications. The breakdown of the BOSS CLASS and SUBCLASS parameters is outlined in \cite{2017MNRAS.468.1156O}.

 \item[5.] IRAC colours: we use the IRAC four-band colour-colour AGN diagnostic proposed by 
\cite{2012ApJ...748..142D} 

\begin{equation}
 {x \ = \ \log_{10}\left(\frac{f_{5.8\mu m}} {f_{3.6\mu m}}\right) , \    \ y = \log_{10}\left(\frac{f_{8.0\mu m}}{f_{4.5\mu m}}\right)}
\end{equation}
\begin{multline}
 x  \ge \ 0.08 \ \wedge \ y   \ge  0.15
\\ 
\wedge  y \ge  (1.21  \times   x)  -  0.27
\\
\wedge y \le  \ (1.21  \times  x)  + 0.27
\\
\wedge  f_{4.5\mu m}   >  f_{3.6\mu m}  >   f_{4.5\mu m}   \wedge   f_{8.0\mu m}   >   f_{5.8\mu m} \\
\label{donley_crit}
\end{multline}


\end{enumerate}

Using these AGN/SFG indicators, we classify the 610 MHz sources as follows:

\begin{enumerate}
  \item[1.] RL AGN: these are sources with redshift information and $L_{\rm 1.4\,GHz} > 10^{25} \ {\rm W\,Hz^{-1}}$ or $q_{24\mu {\rm m}}$ below the M82 locus.
  
  \item[2.] RQ AGN: sources with redshifts and above the M82 locus threshold for selecting RL AGN. Furthermore, these sources are classified as AGN by at least one of the AGN diagnostics listed above.
  
  \item[3.] SFG: these are sources with redshifts
  that do not show evidence of AGN activity in any of the diagnostics.
  
  \item[4.] Unknown: these sources include those that are unmatched to SERVS IRAC12 positions, sources with no redshift and sources with redshift but no information on the parameters needed for the AGN classification. 
\end{enumerate}

\begin{figure}
 \centering
 \centerline{\includegraphics[width = 0.53\textwidth]{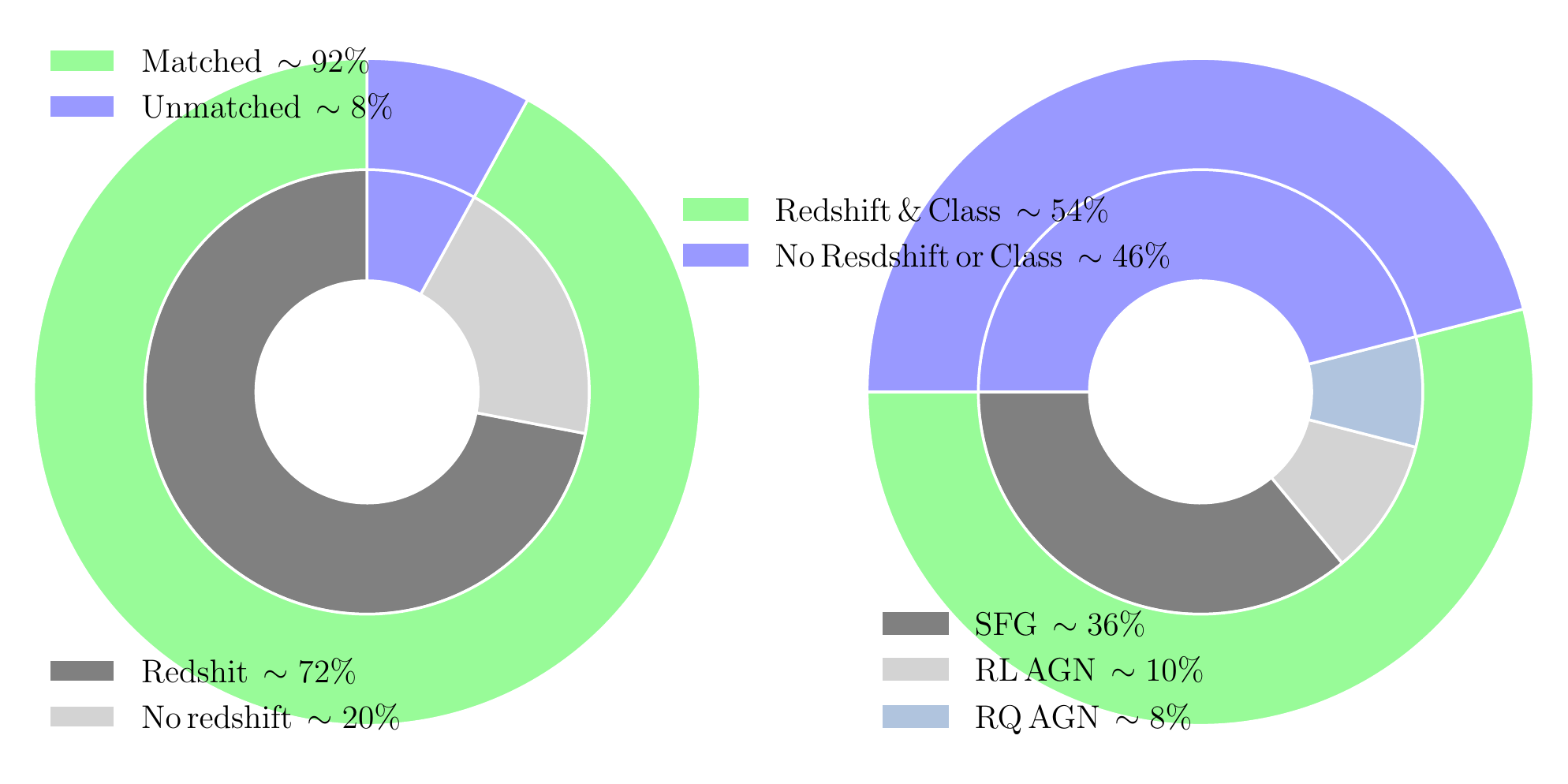}}
\caption{Top: Nested donut chart with two groups illustrating the GMRT sample that is matched (92\%, green) and unmatched (8\%, violet) to SERVS/UKIDSS positions. The three subgroups represents the GMRT sample with redshift (72\%, dark grey), no redshift (20\%,light grey) and the unmatched (8\%, violet). Bottom: Nested donut chart with two groups illustrating the GMRT sample with redshift and at least one AGN indicator (54\%, green), redshift and no AGN indicator (46\%, violet). The four subgroups represents the fraction classified as SFG (36\%, dark grey), RL AGN (10\%,light grey), RQ AGN (8\%, violet) and redshift and no AGN indicator (46\%, violet).}
\label{source_class} 
 \end{figure}

The top panel of Figure~\ref{source_class} shows 
a nested donut chart with two groups illustrating the GMRT sources matched (92\%, lime) and unmatched (8\%, violet) to SERVS/UKIDSS positions. The three subgroups represents the GMRT sources with redshift  (72\%, dark grey), no redshift  (20\%,light grey) and unmatched  (8\% ,violet).
The bottom panel represents a nested donut chart with two groups illustrating the GMRT sources with redshift and AGN classification possible for at least one AGN diagnostics (54\%, green), redshift but no AGN classification possible (46\%, violet). The four subgroups represents the fraction classified as SFG (36\%, dark grey), RL AGN (10\%, light grey), RQ AGN (8\%, violet) and redshift and no AGN classification possible (46\%, violet).

\begin{table*}
\caption{Total number of SFGs, RQ AGNs and RL AGNs from the selection criteria.}
\centering
\begin{tabular}{|c|c|c|c|}
\hline
\hline
Class & Number & Fraction ($\%$) & Fraction ($\%$)\\
  & & (Sub-Sample with Redshift and Class) & (Full Sample)\\
\hline
SFG    & 1685         & 73$\%$ & 39$\%$ \\
RQ AGN &  281         & 12$\%$ &  7$\%$ \\
RL AGN &  338         & 15$\%$ &  8$\%$ \\
No Redshift or Class  & 1986 & - & 46$\%$ \\
\hline
\end{tabular}
\label{classification.tab} 
\end{table*} 

\begin{figure*}
\centering
\includegraphics[width = 0.49\textwidth]{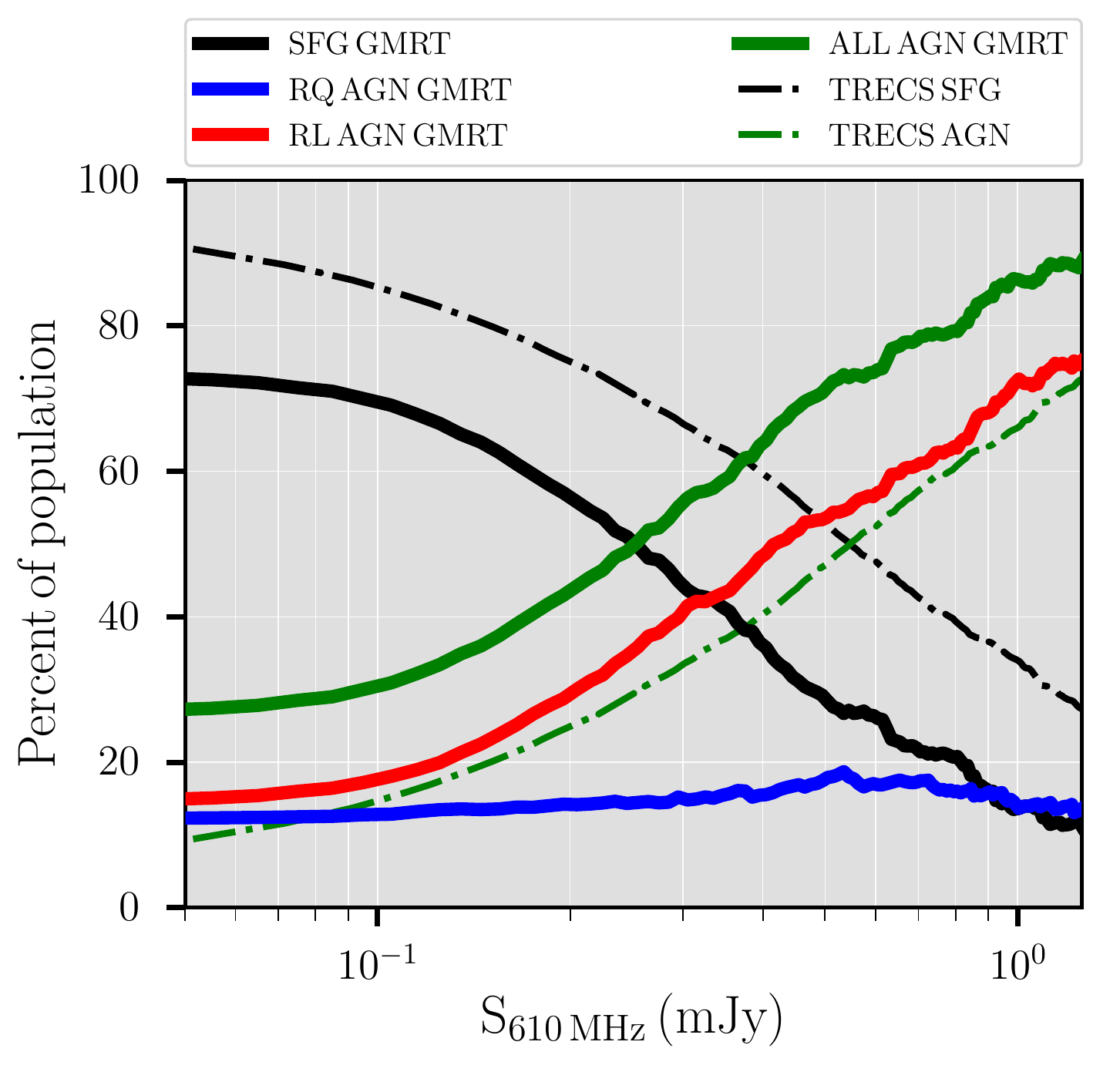}
\includegraphics[width = 0.49\textwidth]{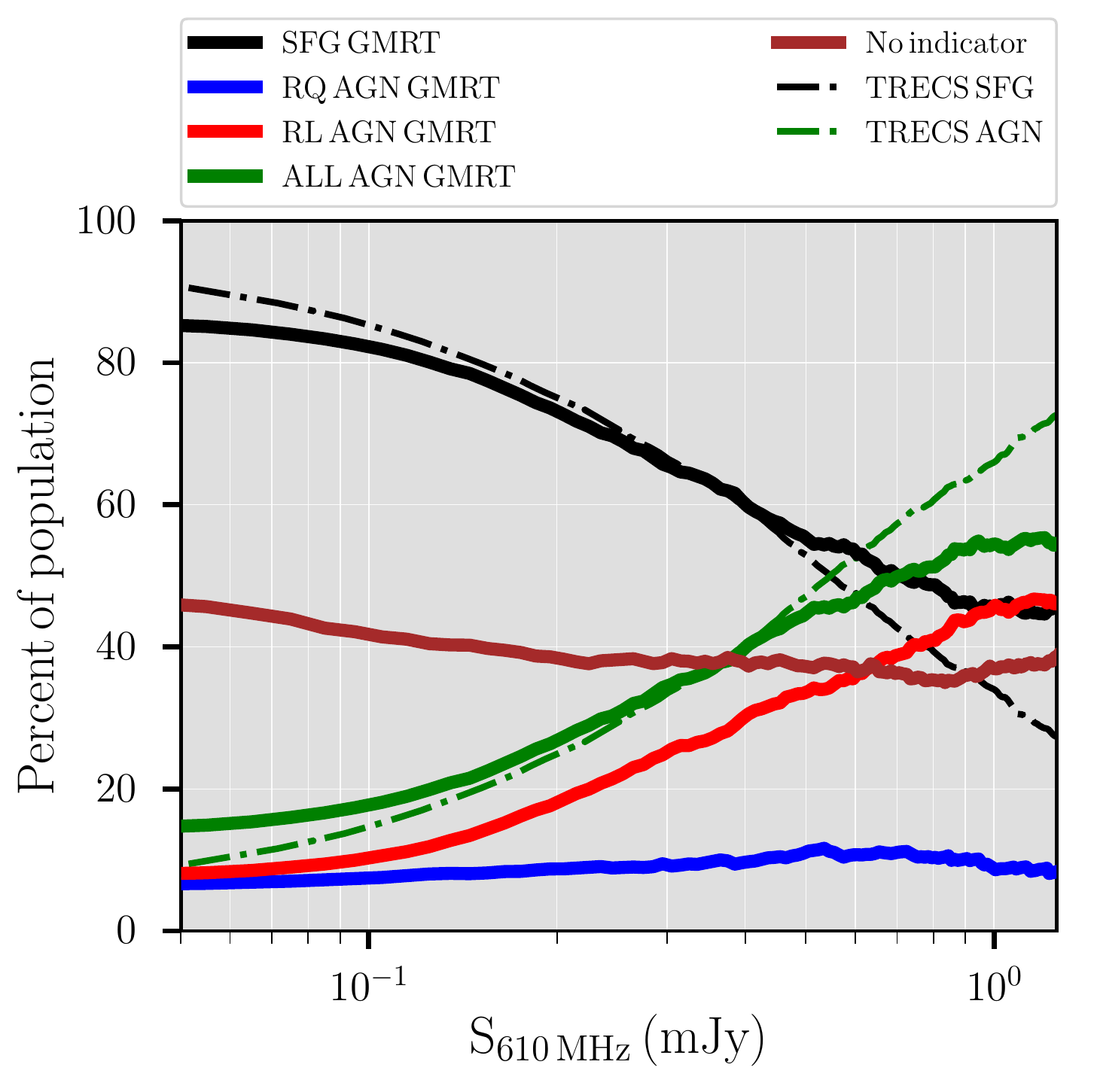}
\caption{The relative fraction of AGNs and SFGs as a function of minimum flux density. The solid lines show the fraction of the sample with a flux density greater than $S_{\rm 610\,MHz}$ which is classified as RQ\,AGN (blue), RL\,AGN (red), all AGN (green) and SFG (black). The dashed dotted green and black lines in both panels represents the relative fraction of AGN and SFG predicted by T-RECS \citep{Bonaldi2019}.}
\label{fig:fraction} 
\end{figure*}

 The substantial number of objects in our sample allows us to study how the faint radio source population changes with flux density.
The left panel of Figure~\ref{fig:fraction} shows the fraction of objects in each class in our sample as a function of limiting flux density.  For a given flux density, $S_{\rm 610\,MHz}$, the plot shows the fraction of objects that are classified as SFG, RL\,AGN, and RQ\,AGN in the sample of objects above that flux density. The green curves show the fraction for the total AGN population. The curves highlight the dramatic change in population over this flux density range. The SFG fraction exhibits a monotonic increase with decreasing flux density from $\rm{\sim10\%\,to\, 72\%}$. RL AGNs decrease rapidly from being the dominant population above $\rm{\sim\,1\,mJy}$ to the smallest fraction below $\rm{\sim\,0.3\,mJy}$. The fraction of RQ AGNs remains roughly constant with flux density just above $\rm{\sim10\%}$. Above $\rm{\sim0.7\,mJy}$, the fraction of RQ AGNs is higher than that of SFGs. \cite{2015MNRAS.452.1263P} identied 626 radio sources with redshifts and classified 55\%, 25\% and 20\% as SFGs, RQ AGNs and RL AGNs respectively from a deep 1.4 GHz sample reaching a $\rm{32.5\,\mu}$Jy flux limit over $\rm{0.29\,deg^{2}}$ of the ECDFS VLA image. They further confirmed the main results of \cite{2011ApJ...740...20P} that AGNs dominate at large flux densities ($\gtrsim$ 1 mJy) but SFGs become the dominant population
below $\approx$ 0.1 mJy. 
\citet{2013MNRAS.436.3759B}  reported that SFGs represent $\rm{57\,\pm\,3\%}$ of the sub-millijansky sample, are missing at high flux densities but become the dominant population below $\rm{\approx\,0.1\,mJy}$, reaching 61\% at the
survey limit. Radio-quiet AGNs represent $\rm{26\pm6\%}$ (or 60\% of all AGNs) of sub-millijansky sources but their fraction appears to increase at lower flux densities, where they make up 73\% of all AGN and $\rm{\approx\,30\,\%}$ of all sources at the survey limit, up from $\rm{\approx\,6\%}$ at $\rm{\approx\,1\,mJy}$.  These results from previous observations are in good agreement with what we report. The fact that we find more SFGs at faint flux densities can be attributed to our survey going deeper than previous surveys. We compare our results to the relative fraction of AGN and SFG computed for T-RECS by \cite{Bonaldi2019} (see dashed dotted green and black lines in both panels of Figure~\ref{fig:fraction}) and find that the fraction of our classified sources do not agree with T-RECS. When we add the fraction of sources that have no classification to the SFGs fraction, we see that below $\rm{\sim0.6\,mJy}$ our computed AGN and SFGs fraction agrees well with T-RECS (see the right panel in Figure~\ref{fig:fraction}).
Table~\ref{classification.tab} presents the total number of AGNs (including RL and RQ AGNs) and SFGs with respect to sources with redshifts and AGN classification possible as well as the full GMRT sample.

\section{Multi-Frequency Radio Spectral Indices}\label{spectral.sec}
Radio spectral energy distributions (SEDs) provide useful information that can be used to differentiate between 
sources types according to their dominant emission mechanisms \citep{2014MNRAS.439.1556M}.
We computed the spectral index between 325 - 610 MHz, 610 - 1400 MHz and 610 - 5000 MHz for our EN1 sample using the GMRT 325 MHz deep survey by \citet{2009MNRAS.395..269S}, 
the Faint Images of the Radio Sky at Twenty Centimeters (FIRST) 1400 MHz survey \citep{1995ApJ...450..559B} and the JVLA 5000 MHz
Ultra Deep Survey by \citet{2014ASInC..13...99T}. Since the images mentioned above have different resolutions (see Table~\ref{tab_multi-matches}), special care must be taken when we analyse results based on different frequency-selected samples.

From the commonly used simple power law model, a negative $\rm{\alpha}$ is indicative of sources dominated by synchrotron emission, such as radio galaxies. An $\rm{\alpha\,\sim\,0}$ may indicate either a star-forming galaxy  dominated by free-free emission optically thin or optically thick synchrotron emission in core-dominated AGNs.
Inverted $\rm{\alpha\,>\,0}$  spectra in the GHz regime can be associated to very young compact sources (Gigahertz Peaked Sources, GPS) or to Advection-Dominated Accretion Flow (ADAF) sources (see e.g. \citealt{1994ApJ...428L..13N}). Thus, radio spectra are useful in unveiling the physical processes in radio sources \citep{2010A&A...510A..42P,doi:10.1093/mnras/sty1818}.

\subsection{Radio Spectral Index vs Flux Analysis}
\begin{figure}
 \centering
 \centerline{\includegraphics[width = 0.5\textwidth]{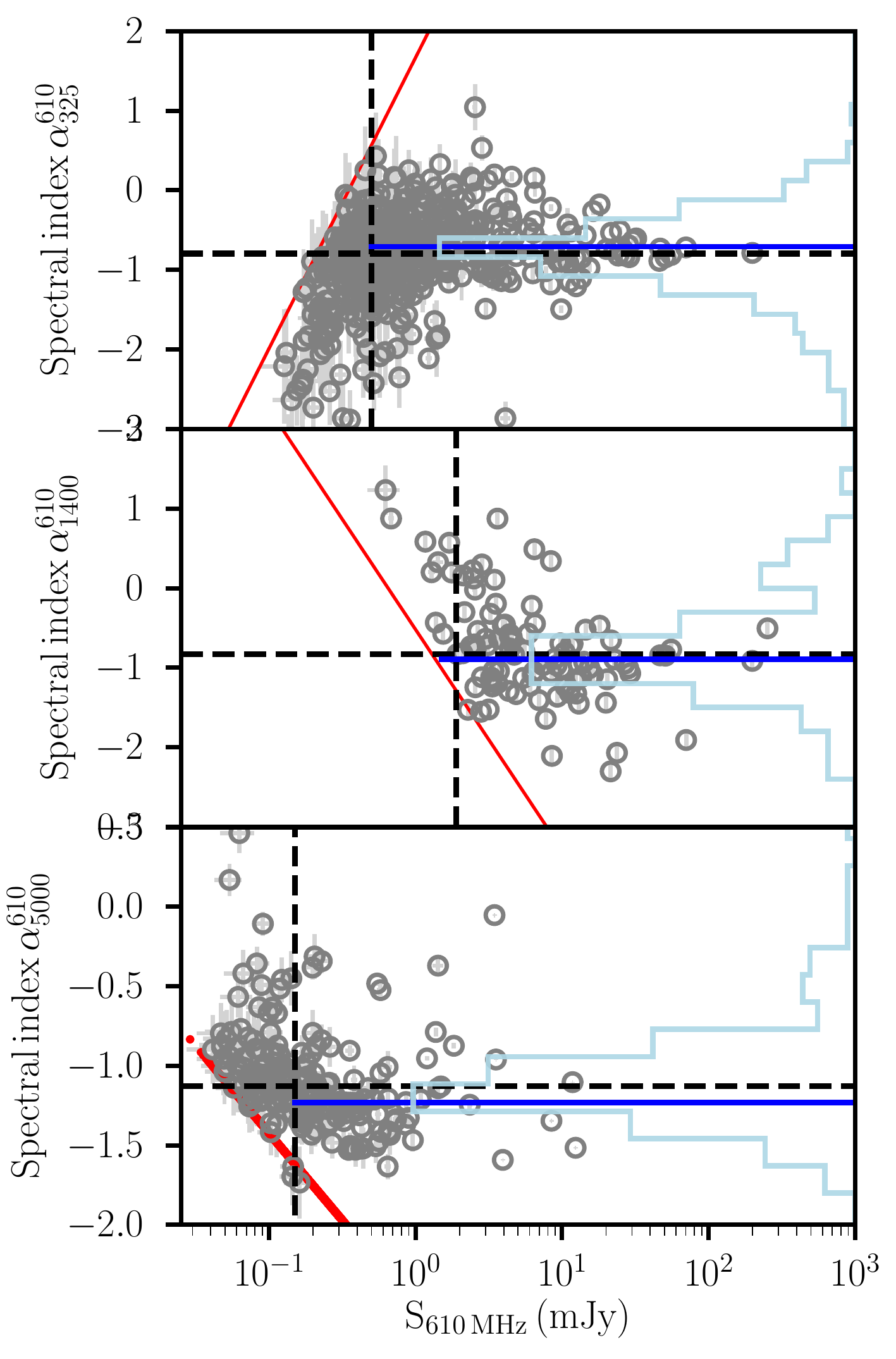}}
\caption{colour-flux diagrams comparing the 325-610 MHz (top), 1400-
610 MHz (middle), and 5000-610 MHz (bottom) spectral indices for EN1 GMRT  Deep 610 MHz
cross-identified sources. The solid red sloping line in each panel marks the flux density traced by the nominal
detection limit of 325, 1400 and 5000 MHz respectively(see Table~\ref{tab_multi-matches}). The distribution of the spectral index between each frequency is shown as blue histogram in each panel.  The dashed vertical lines in each panel represents the flux limit we impose when restricting our statistical analyses to a much brighter sub-sample. The blue horizontal lines in each panel represents the median spectral indices we measure for sources above the imposed flux limits.} 
\label{fig:specindx_vs_flux} 
 \end{figure}

We investigate the spectral index properties of 610 MHz low-frequency selected sources. 
We estimated the median 
and the error on the median using the median absolute deviation estimator as this is 
a more robust measure of the variability of a univariate sample of quantitative data than the standard deviation
 \citep{doi:10.1080/01621459.1993.10476408}.
Table~\ref{tab_multi-matches} summarizes the number of matches between the 610 MHz catalogue and the samples at other frequency. Figure~\ref{fig:specindx_vs_flux} shows the 610 - 325 MHz, 610 - 1400 MHz and 610 - 5000 MHz colour-flux diagrams. The distribution of the spectral index between each frequency pair is shown as blue histogram in each panel. For the top panel of Figure~\ref{fig:specindx_vs_flux}, we note that only 479 of our GMRT  610 MHz sources have a counterpart at 325 MHz, therefore, $\mathrm{\alpha^{610}_{325}}$ estimates are available only for 13 per cent of our radio-detected sources at 610 MHz. We find that $\mathrm{\alpha^{610}_{325}}$ estimates range from -2.7 to 1.8 with a median value of $\mathrm{-0.80\,\pm\,0.29}$. \cite{2009MNRAS.395..269S} reported a median spectral index between $\mathrm{\alpha^{610}_{325}}$ 1.28 from 325 MHz studies of EN1 using the GMRT. They attributed their median value to an extra contribution of exceedingly steep diffuse emission being detected at the lower frequency.
In the middle panel, only 99/4290 ($\rm{\sim}$ 2.3\%) of our 610 MHz detected  sources have counterparts at 1.4 GHz.
The {$\mathrm{\alpha^{610}_{1400}}$} estimates range from -2.5 to 1.1 with a median value of $\mathrm{-0.83\,\pm\,0.31}$. The bottom panel has the second highest number of sources since the JVLA 5000 MHz  Deep only covers an area of $\rm{0.12\,deg^{2}}$ (see Table~\ref{tab_multi-matches} ) and this is only a small region of the 610 MHz image. Only 204/4290 ($\rm{\sim}$ 4.8\%) of our 610 MHz sources have counterparts at 5 GHz with a median value of $\mathrm{-1.12\,\pm\,0.15}$.

 At the lowest fluxes we are only  sensitive to increasingly steeper (top  and middle panels) or flatter sources (bottom panel), hence the median values that we derive can  be biased and unreliable. We therefore restrict our statistical analyses to a much brighter sub-sample and measure the median spectral index, where the red lines (see Figure~\ref{fig:specindx_vs_flux})  are not biasing too much the median spectral indices. We find that the median spectral index for $\mathrm{\alpha^{610}_{325}}$ for a flux range corresponding to $\rm{S_{610\,MHz}}$ > 0.5 mJy represented by the vertical black dash line in the top panel of Figure~\ref{fig:specindx_vs_flux} is $\mathrm{-0.71\pm0.27}$ (see the horizontal blue solid line). In the middle panel, we measure a median spectral index for $\mathrm{\alpha^{610}_{1400}}$ over a flux range corresponding to $\rm{S_{610\,MHz}}$ > 1.9 mJy to be $\mathrm{-0.89\pm0.28}$ (see the horizontal blue solid line). For the bottom panel, we measure a median spectral index for $\mathrm{\alpha^{610}_{5000}}$ over a flux range corresponding to $\rm{S_{610\,MHz}}$ > 0.15 mJy to be $\rm{-1.23\pm0.12}$ represented by the horizontal blue solid line.

 The median spectral index  $\mathrm{\alpha^{610}_{325}}$ as a function of flux densities > 0.5 mJy is found to be approximately -0.71 from the original -0.80. This can be attributed to the fact that above the imposed flux density limit, most of the sources we select are AGNs and thus reduces the median spectral index we measure. The median spectral index $\mathrm{\alpha^{610}_{1400}}$ is found to be approximately -0.83
to -0.89, based on an almost unbiased sources with flux densities > 1.9 mJy. Statistical analyses of \cite{2009MNRAS.397..281I} showed no clear evolution for the median spectral index, $\mathrm{\alpha^{610}_{1400}}$, as a function of flux density and that $\mathrm{\alpha^{610}_{1400}}$
 was found to be approximately -0.6 to -0.7 based on an almost unbiased 10$\sigma$ criterion, down to a flux level of $\mathrm{S_{1.4GHz}\,\gtrsim\,100\,\mu Jy}$. \cite{1974A&A....35..393K} found from a small sample of sources
with $\rm{S_{610MHz}\,\gtrsim\, 10mJy}$ a spectral index distribution of $\mathrm{\alpha^{610}_{1400}\,=\,-0.52\pm0.39}$ using the Westerbork Synthesis Radio Telescope (WSRT). With respect to higher frequency surveys with a broad distribution, this was an unusual result. However, \cite{1979A&A....73..107K} using a much larger sample presented similar result of $\mathrm{\alpha^{610}_{1400}\,=\,-0.68\pm0.31}$.  A detailed spectral index analysis using the other available radio data and adding the upper/lower  limits for each source to get reliable estimates of median spectral indices through survival analysis is deferred to later works.

For the remaining 419/901 ($\sim$ 47 per cent, for 325 MHz), 688/818 ($\sim$ 84 per cent, for 1400 MHz), 170/330 ($\sim$ 52 per cent, for 5000 MHz) detected with no counterparts in the GMRT 610 MHz we derive the nominal detection limit in 325, 1400 and 5000 MHz (see Table~\ref{tab_multi-matches}) respectively represented by the solid red line in each panel.

\begin{table*}
 \centering
 \caption{The radio surveys that were used to form the multi-frequency samples through cross-matching with the 610 MHz catalogue.}
 \begin{tabular}{|c|c|c|c|c|c|c|}
 \hline
 \hline
Survey& Frequency  &Resolution    & Area covered & rms  & Number of Sources &Number of matches \\
  &(MHz)&(arcsec)&($\rm{deg^2}$)&($\rm{\mu Jy}$)&\\
 \hline
EN1 GMRT Deep &  325 & 10  &  1.5  & 70 & 901 & 479 \\
VLA FIRST (All Sky) & 1400 &  5   & --    & 150 & -- & 99 \\
EN1 JVLA Deep & 5000 &  2.5 &  0.12 &  1 & 387 & 204 \\
 \hline
 \end{tabular}
 \label{tab_multi-matches} 
 \end{table*}

\subsection{Radio colour-colour plot}
\begin{figure*}
 \centering
 \centerline{\includegraphics[width = 0.8
\textwidth]{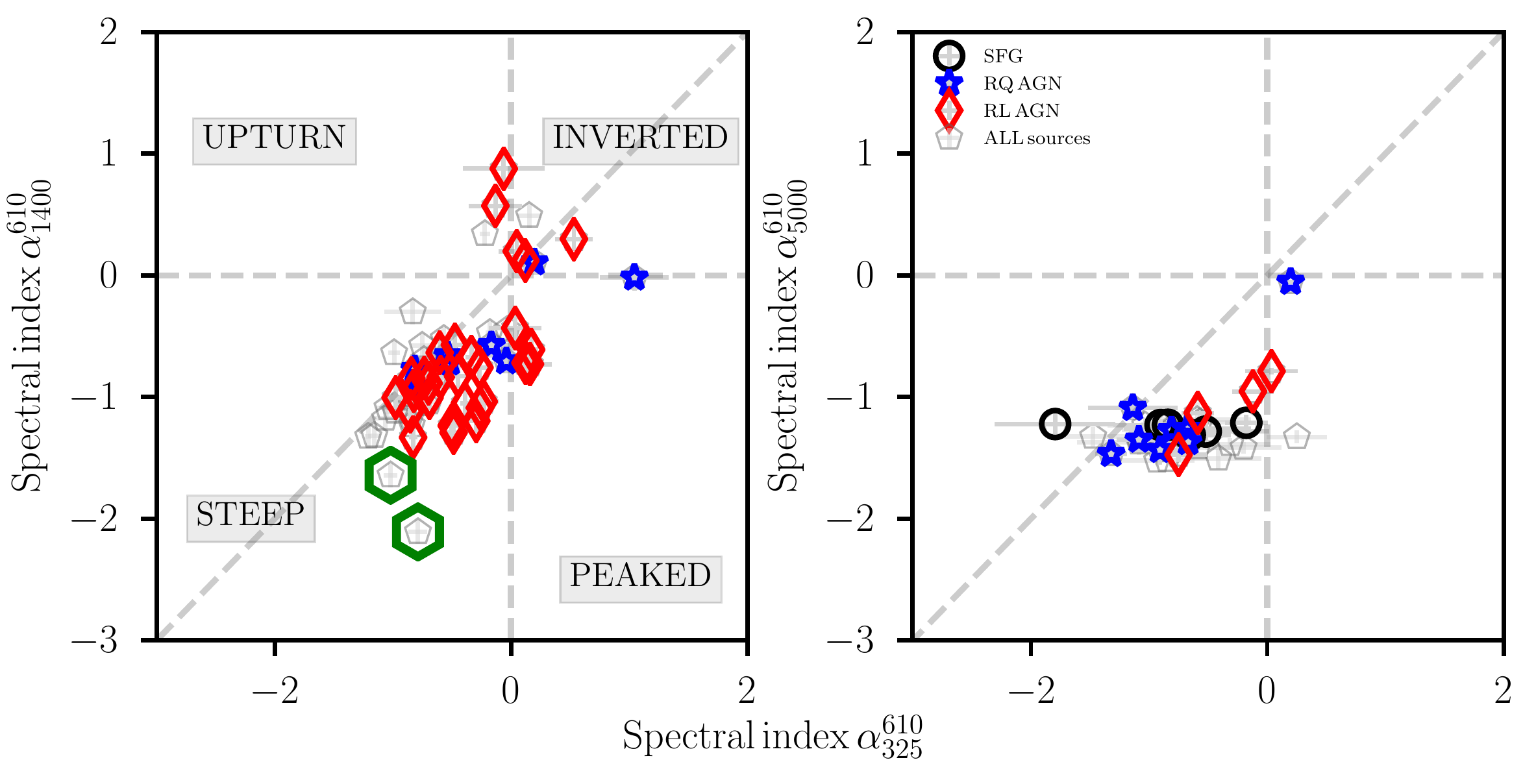}}
\caption{Radio colour - colour plots for sources the EN1 GMRT 610MHz Deep sample. Left: the spectral indices of the sample between 325 and 610 MHz against
the spectral indices between 1400 and 610 MHz.
Right: the spectral indices of the sample between 325 and 610 MHz against the spectral indices between 5000 and 610 MHz. SFG, RQ AGN, RL AGN and all sources  indicator are represented with black, red, blue and dimgrey colours respectively. The green hexagon in the left panel shows ultra very steep sources (i.e. $\rm{\alpha^{610}_{1400}}$ < -1.5).}
\label{fig:specindx_vs_specindx} 
 \end{figure*}

 Figure~\ref{fig:specindx_vs_specindx} {\bf shows} radio colour - colour plots for the EN1 GMRT 610MHz Deep sample. The spectral indices of the sample between 325 and 610 MHz against the spectral indices between 1400 and 610 MHz (i.e using the spectral indices measure in the middle panel of Figure~\ref{fig:specindx_vs_flux}) is shown in the left panel. The right panel show the spectral indices of the sample between 325 and 610 MHz against the spectral indices between 5000 and 610 MHz. 
We divide the radio colour-colour plot into four quadrants. 
\begin{enumerate}
  \item[1.] Steep and flat spectrum
sources: where($\rm{\alpha^{610}_{1400}}$ $\mid$ $\rm{\alpha^{610}_{5000}}$ $\leq\,0$ $\&$ $\rm{\alpha^{610}_{325}}$ $\leq\,0$)
\item[2.] Peaked spectrum
sources: where ($\rm{\alpha^{610}_{1400}}$ $\mid$ $\rm{\alpha^{610}_{5000}}$ $\leq\,0$ $\&$ $\rm{\alpha^{610}_{325}}$ $>\,0$)
\item[3.] Inverted spectrum
sources: where ($\rm{\alpha^{610}_{1400}}$ $\mid$ $\rm{\alpha^{610}_{5000}}$ $>\,0$ $\&$ $\rm{\alpha^{610}_{325}}$ $>\,0$)
\item[4.] Upturn spectrum sources ($\rm{\alpha^{610}_{1400}}$ $\mid$ $\rm{\alpha^{610}_{5000}}$ $>\,0$  $\&$ $\rm{\alpha^{610}_{325}}$ $\leq\,0$)
\end{enumerate}

It is evident that the majority of our GMRT 610 MHz sources lie in the steep and flat spectrum quadrant in both panels. Moreover, the
scatter around the diagonal line is asymmetric for the first panel (i.e $\rm{\alpha^{610}_{1400}}$ vs $\rm{\alpha^{610}_{325}}$) of Figure~\ref{fig:specindx_vs_specindx} with relatively more number of sources lying below the lower left side of the diagonal line (i.e at  $\rm{\alpha^{610}_{1400}}$ <  $\rm{\alpha^{610}_{325}}$ ), indicating a steepening of the spectrum at higher frequencies (despite the fact the $\rm{\alpha^{610}_{1400}}$ MHz sample is biased towards flat spectrum sources. This is consistent with the RL AGN classification of most such sources). 

The right panel show a  slight scatter around the diagonal. It is interesting to note that  sources lying below the diagonal line (i.e. $\rm{\alpha^{610}_{325}}$ < $\rm{\alpha^{610}_{5000}}$) are mostly SFGs or RQ AGNs. This sample has more SFGs as it is created from the deep, narrow 5000 MHz survey (see Table~\ref{tab_multi-matches}). 
In the first case the flattening of the spectrum going at higher frequency may be due to an increase of the free-free emission contribution at high frequency; in the case of RQ AGN the flattening may be due to the emergence of core-dominated emission.


\subsection{SFG and AGN spectral indices}
\begin{figure*}
 \centering
 \centerline{\includegraphics[width = 0.85\textwidth]{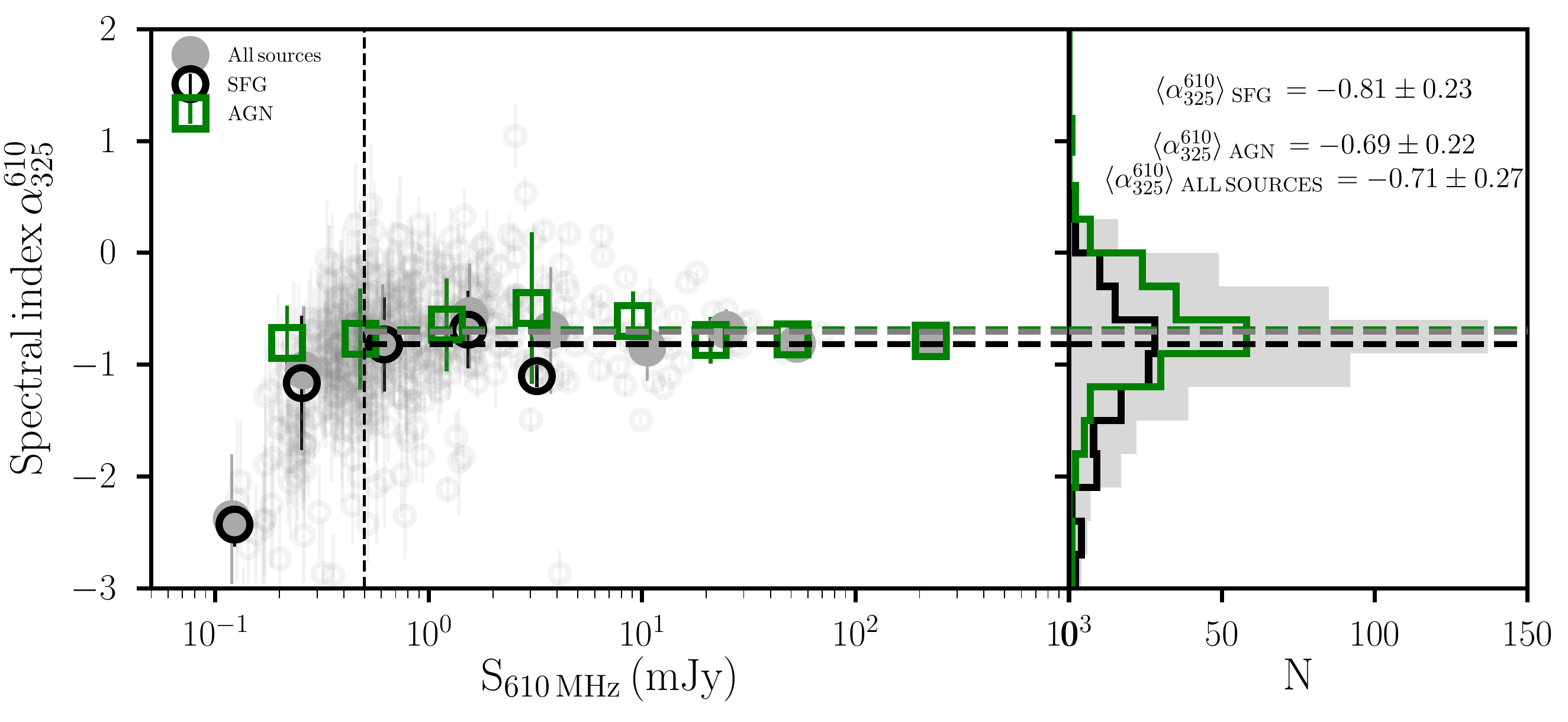}}
 \centerline{\includegraphics[width = 0.85\textwidth]{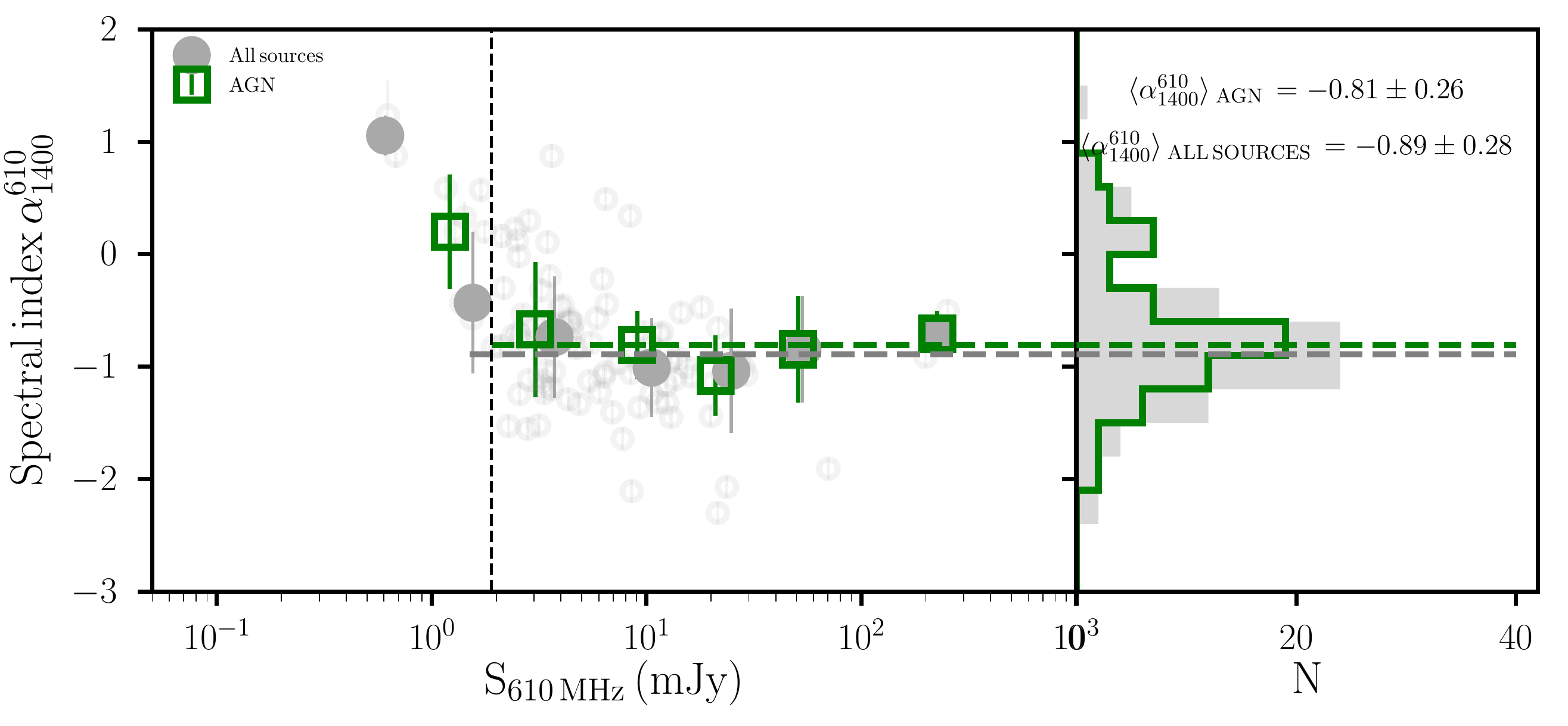}}
\caption{Top panel: The distribution for $\rm{\alpha^{610}_{325}}$ is computed over a flux range corresponding to $\rm{S_{610\,MHz}}$ > 0.5 mJy. The filled grey circles, open black circles and open green squares represents all sources, SFGs and AGNs respectively in logarithmic bins of 0.4.
Bottom panel: The distribution for $\rm{\alpha^{610}_{1400}}$ is computed over a flux range corresponding to $\rm{S_{610\,MHz}}$ > 1.9 mJy. The filled grey circles and open green squares represents all sources and AGNs respectively in logarithmic bins of 0.4. The number of SFGs having $\rm{\alpha^{610}_{1400}}$ associations is only one and not included in the plot.}
\label{fig:specindx_610_325_class} 
 \end{figure*}

The distribution of the spectral index between $\rm{\alpha^{610}_{325}}$ and $\rm{\alpha^{610}_{1400}}$ for SFG and AGN is shown as black and green histograms in Figure~\ref{fig:specindx_610_325_class}. The distribution for $\rm{\alpha^{610}_{325}}$ is computed over a flux range corresponding to $\rm{S_{610\,MHz}}$ > 0.5 mJy and that of $\rm{\alpha^{610}_{1400}}$ is computed over a flux range corresponding to $\rm{S_{610\,MHz}}$ > 1.5 mJy  respectively. From Figure~\ref{fig:specindx_610_325_class}, the median and median absolute deviation (MAD) computed over $\rm{S_{610\,MHz}}$ > 0.5 mJy is $\rm{\langle\alpha^{610}_{325}\rangle\,= -0.81\pm0.23}$ for SFGs and $\rm{\langle\alpha^{610}_{325}\rangle\,= -0.69\pm0.22}$ for AGNs. RL and RQ AGNs have a median and MAD of $\rm{\langle\alpha^{610}_{325}\rangle\,= -0.67\pm0.27}$ and $\rm{\langle\alpha^{610}_{325}\rangle\,= -0.71\pm0.22}$ respectively. 

We computed $\rm{\langle\alpha^{610}_{1400}\rangle\,= -0.81\pm0.26}$ over $\rm{S_{610\,MHz}}$ > 1.9 mJy for AGNs (see bottom panel of  Figure~\ref{fig:specindx_610_325_class}). In addition, RL and RQ AGNs have a median and MAD of $\rm{\langle\alpha^{610}_{1400}\rangle\,= -0.89\pm0.28}$ and $\rm{\langle\alpha^{610}_{1400}\rangle\,= -0.68\pm0.10}$ respectively. The number of SFGs having $\rm{\alpha^{610}_{1400}}$ associations is only one and not included in this analysis.
Table~\ref{tab_specindex} shows the breakdown of the number of SFGs, RL AGN, RQ AGN and sources with no classification that have a spectral index.

 \subsection{Ultra steep spectrum sources}
 Ultra steep spectrum sources (USS) radio sources are often associated with radio galaxies at high
redshift (HzRGs $\rm{z \,>\, 2}$) (e.g. see \citealt{1979A&A....80...13B,2008A&ARv..15...67M}). HzRGs are located
in overdense regions in the early Universe and are frequently surrounded by
protoclusters (\citealt{1996yCat..41080079R,1996AAS...189.8302P,1997ApJ...487..644K}).
Studies have shown USS are good candidates for high redshift radio galaxies  \citep{2016MNRAS.462..917R} which are among the most luminous and massive galaxies 
(e.g. \citealt{2005mmgf.conf..374D,2007MNRAS.378..551B,2014A&A...569A..52S}) and are believed to be progenitors of the massive elliptical galaxies in the local
Universe. Their
extremely steep spectrum is generally attributed to radiation losses
of relativistic electrons in the radio lobes, meaning they are most
luminous at lower radio frequencies \citep{2016MNRAS.463.2997M}.
 
In the literature, USS radio
sources are commonly defined as those with spectral index values
$\rm{\alpha\,<\,-1.3}$ (\citealt{10.1093/mnrasl/slt008,2016A&A...593A.130H}). We selected sources that had spectral indices between 610 MHz and 1.4 GHz steeper than $\rm{\alpha\frac{610}{1400}\,=\,-1.2}$. Using this criterion, we find two sources with ultra-steep spectra in the GMRT 610 MHz sample. The 610 - 325 MHz spectral index for the first USS candidate (i.e source with GMRT ID 713) is $\rm{\alpha^{610}_{325}\,=\,-1.01}$ whereas the 610 - 1400 MHz spectral index is $\rm{\alpha^{610}_{1400}\,=\,-1.65}$. The second USS candidate (i.e source with GMRT ID 2388) has a 610 - 325 MHz spectral index of $\rm{\alpha^{610}_{325}\,=\,-0.79}$ and  a 610 - 1400 MHz spectral index of $\rm{\alpha^{610}_{1400}\,=\,-2.10}$. This source is at the detection of threshold of the NRAO VLA Sky Survey (NVSS) \citep{1998AJ....115.1693C} and has flux density of 2.1$\pm$0.4 mJy at 1.4 GHz. If we use this flux, then 610 - 1400 MHz spectral index is  $\rm{\alpha^{610}_{1400}\,=\,-1.6}$. 
Both candidate USS sources do not have redshift associations in our catalogue, hence these sources are more likely to be HzRGs.

\begin{figure*}
 \centering
 \centerline{\includegraphics[width = 0.85\textwidth]{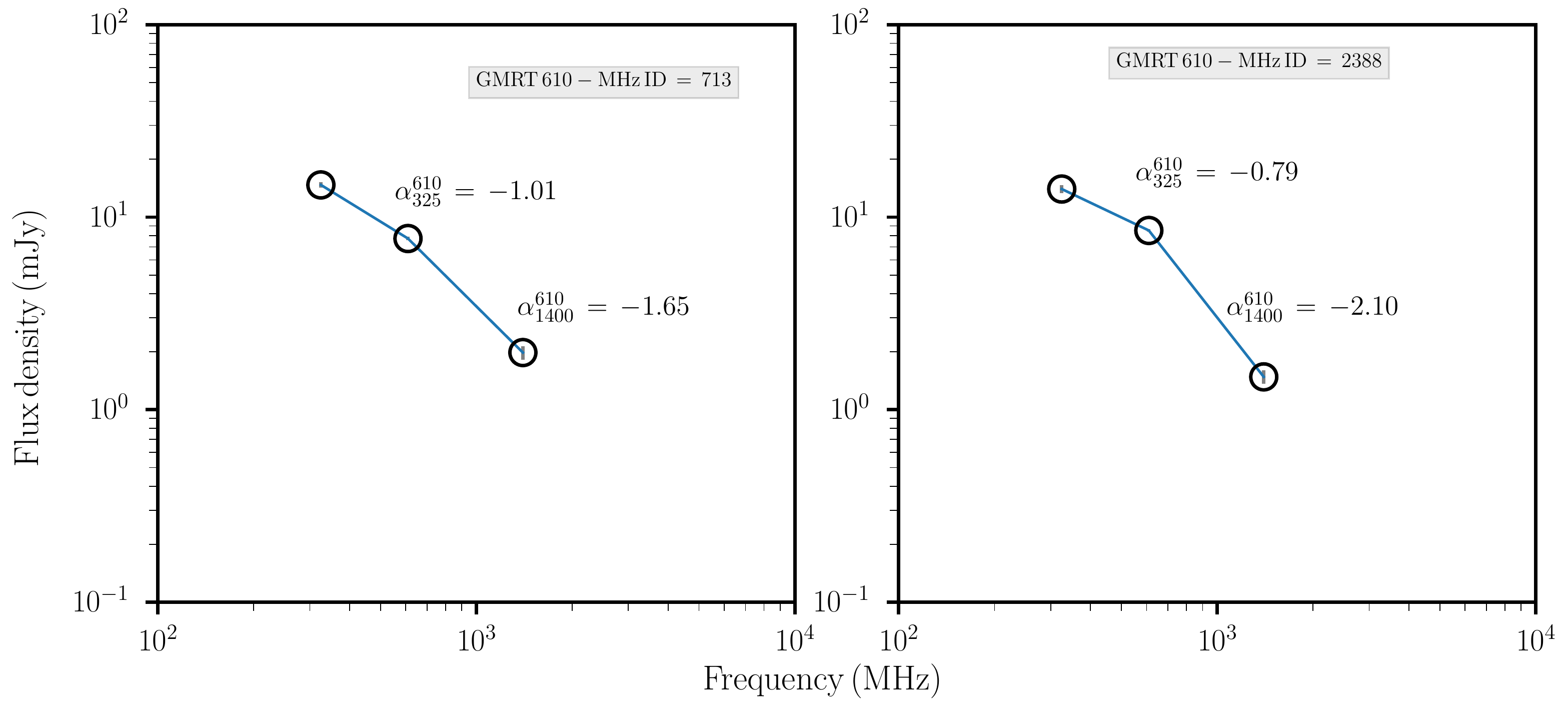}}
\caption{Flux density as a function of frequency for the two ultra steep spectrum sources identified in Figure~\ref{fig:specindx_vs_specindx}. The radio SED for the two USS is from 325 MHz to 1400 MHz. Also shown are the spectral indices measured between data
available for this source at various radio frequencies. }
\label{fig:specindx_steep_spec} 
 \end{figure*}

\begin{figure*}
 \centering
 \centerline{\includegraphics[width = 0.98\textwidth]{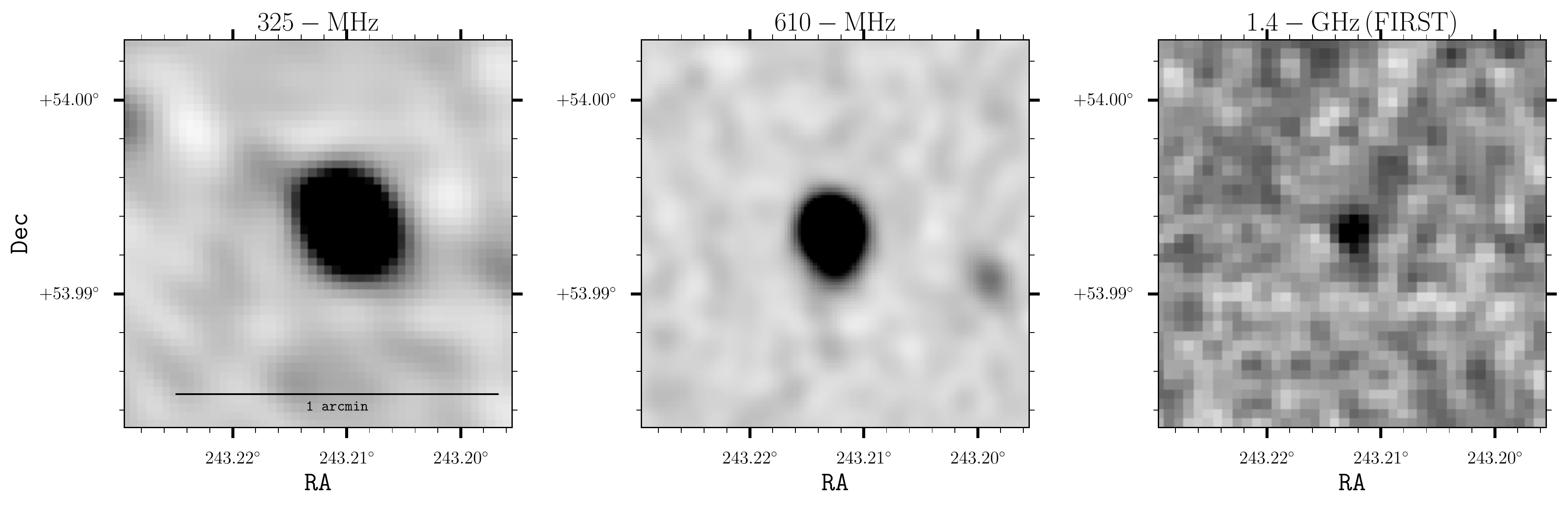}}
 \centerline{\includegraphics[width = 0.98\textwidth]{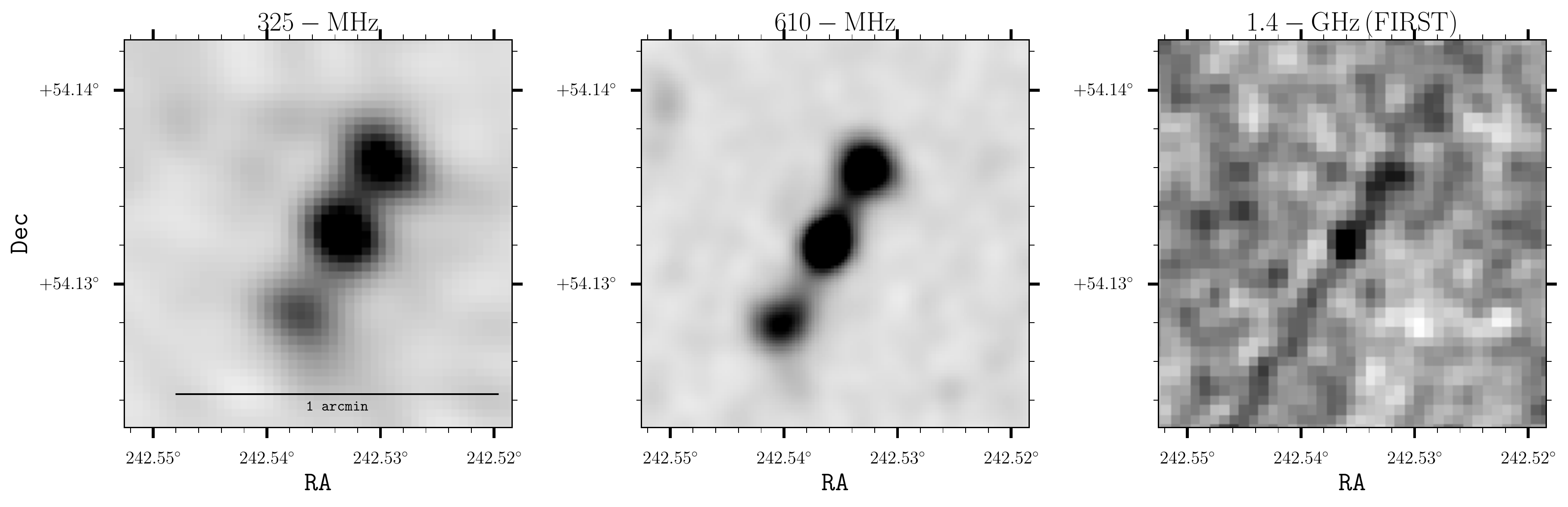}}
\caption{Candidate USS radio sources at three different frequencies from 325 MHz  , 610 MHz and 1.4 GHz (FIRST) respectively for the two sources.}
\label{USS_postage.fig} 
 \end{figure*}

Figure~\ref{fig:specindx_steep_spec} shows the flux density as a function of frequency for the two very steep spectrum sources identified in Figure~\ref{fig:specindx_vs_specindx}, the two-point spectral index values are also printed on the Figure. Postage stamp images of these sources are presented in Figure~\ref{USS_postage.fig}.

\begin{table}
 \centering
 \caption{Number of SFGs, RL AGN, RQ AGN and sources with no classification that have a spectral index.}
 \begin{tabular}{|c|c|c|c|c|}
 \hline
 \hline
$\rm{\alpha}$& SFG  &RL AGN    & RQ AGN&No classification \\
 \hline
$\rm{\alpha^{610}_{325}}$ &  122 & 125   &  48&184 \\
$\rm{\alpha^{610}_{1400}}$ &  1   & 53& 10&35 \\
$\rm{\alpha^{610}_{5000}}$& 73 & 12 &23  &96 \\
 \hline
 \end{tabular}
 \label{tab_specindex} 
 \end{table}

\section{Summary and Conclusions}\label{sum.sec}
We report deep 610 MHz GMRT observations of the EN1
field, a region of $\rm{1.864\,deg^{2}}$. We achieve a nominal sensitivity of $\rm{7.1 \mu\,Jy\, beam^{-1}}$. From our 610 MHz mosaic image, we recover 4290 sources after accounting for multiple component sources down to a $\rm{5\sigma}$ flux density limit of $\rm{35.5\,\mu\,Jy}$.

From this data, we derive the 610 MHz source counts applying corrections for completeness, resolution bias and Eddington bias. The counts are within the scatter of most previous source counts from other surveys at 610 MHz and with extrapolated models of the low-frequency source population; the most obvious exception is the \citet{2008MNRAS.388.1335W} source counts. The counts show a flattening below $\sim\,1$ mJy as a result of the increasing contribution of SFGs (\citealt{2004MNRAS.355L...9R,2015MNRAS.452.1263P,Padovani2016}).

Our radio catalogue was cross-matched against SERVS, UKIDSS and other multi-wavelength datasets. Using the different radio, mid-infrared, optical and X-ray AGN indicators explored in \citet{2017MNRAS.468.1156O}, we have efficiently separated the radio source population with redshift into three classes: SFGs, RQ AGNs and RL AGNs. The relative contribution of the three classes of sources to the subsample of radio sources with redshifts and at least one multi-wavelength diagnostic is as follows: $\sim$73\% SFGs, $\sim$12\% RQ AGNs and $\sim$15\% RL AGNs. Compared to our previous analysis over a smaller area at 610 MHz in the same field, our results indicate a continued increase in the relative fraction of SFGs with decreasing flux density. \citet{2017MNRAS.468.1156O} reported that RQ AGNs dominate the AGN population but in this work spanning a larger area of the same field we conclude that RL AGNs actually dominate. The significantly higher fraction of SFGs in our sample may also partially arise from the selection at lower frequency, where at a given flux density threshold flat-spectrum AGN cores are preferentially detected at 1.4 GHz. 

 We matched our 610 MHz catalogue and compared with catalogues from other surveys at different frequencies. In this regard, we form a sample with which to study the spectral index properties of low-frequency radio sources. We measure the median spectral index between 610 - 325 MHz, 610 - 1400 MHz and 610 - 5000 MHz. Our sample is dominated by steep-spectrum sources as expected for low-frequency selected sources. We measure a median spectral index of $\mathrm{\alpha^{610}_{325}\,=\,-0.80\,\pm\,0.29}$, $\mathrm{\alpha^{610}_{1400}\,=\,-0.83\,\pm\,0.31}$ and $\mathrm{\alpha^{610}_{5000}\,=\,-1.12\,\pm\,0.15}$. We note that the median spectral index we measure at other frequencies for our sample is currently severely limited by the sensitivity of the high-frequency reference.  The radio colour-colour plot (i.e $\rm{\alpha^{610}_{1400}}$ vs $\rm{\alpha^{610}_{325}}$) reveals a  steepening which is consistent with our RL AGN classification. \citet{2011MNRAS.412..318M} showed that a spectral index of $\rm{\alpha\,=\,0.5}$ provides  a  clean way of distinguishing flat-spectrum/compact sources from steep-spectrum/extended sources. These extended sources emit synchrotron
radiation at relatively high frequencies where they are optically thin, implying the existence of fast electrons moving in a magnetic field which is a signature of RL AGN (see, \citet{Padovani2016}). Thus, the  steepening of our RL AGN sources can be attributed to the systematic increase in the synchrotron age of the relativistic jets extending well beyond
the host galaxy, i.e. an increase from the lobes' head towards their flaring ends (see, \citealt{1989MNRAS.240..591S,2010A&A...510A..84M}). 

Restricting our statistical analyses to a much brighter sub-sample, $\rm{S_{610\,MHz}}$ > 0.5 mJy for $\rm{\alpha^{610}_{325}}$ and $\rm{S_{610\,MHz}}$ > 1.9 mJy for $\rm{\alpha^{610}_{1400}}$, we measure a -0.71 and -0.89 respectively. The median spectral indices between 610 - 325 MHz of the bright sample for SFGs and AGNs is $\rm{-0.81\pm0.23}$ and $\rm{-0.69\pm0.22}$ respectively. 
We also measure a median spectral index between 610 - 1400 MHz of $\rm{-0.81\pm0.26}$ for AGNs  over $\rm{S_{610\,MHz}}$ > 1.9 mJy.

By adopting the definition of a USS object as a radio source with $\rm{\alpha\, <\,-1.3}$, we find a total of two USS radio sources. The two candidate USS sources have no corresponding redshift association (both spectroscopic and photometric) from the multi-wavelength catalogue (see Section~\ref{crossmatch.sec} and Section~\ref{redshift.sec}) and therefore remain unclassified. \citet{2018MNRAS.475.5041S} defined a sample of USS radio sources from the TGSS ADR1 at 150 MHz to search HzRGs.  They used the TGSS along with FIRST and NVSS at 1.4 GHz to select sources with spectral indices steeper than -1.3 resulting in a final sample  consisting of 32 sources.  Currently, most powerful distant radio galaxy is at $\rm{z\,=\,5.7}$ \citep{2018MNRAS.480.2733S} with an ultra-steep spectral index, $\rm{\alpha^{150\,MHz}_{1.4\,GHz}\,=\,-1.4}$ (see \citealt{2018MNRAS.475.5041S}). Although we have no redshift estimates for our two candidate USS sources, chances of them being HzRGs is very high. However, there is also the possibility of their being dust obscured radio AGNs at lower redshifts, low luminosity RQ AGNs which are pretty much indistinguishable from SFGs in terms of the radio emission at lower redshifts  present in our sample. Follow-up observations are essential to confirm that the two USS sources are HzRGs.

 A detailed and more complete analysis of the evolutionary properties of the different classes of sources in our GMRT sample, in comparison with other observational and modeling work will be the subject of forthcoming papers.
Upcoming large radio continuum surveys with the SKA pathfinders and precursors \citep{2013PASA...30...20N}, such as the MeerKAT International GHz Tiered Extragalactic Exploration (MIGHTEE) Survey \citep{2016mks..confE...6J} with MeerKAT \citep{2016mks..confE...1J}, will  detect  millions  of  radio  sources down to fainter flux limits than we explored in this  paper.  It  is  therefore  extremely  important  to  be  able  to predict  which  kind  of  sources  these  facilities  will  observe  and  what  are  the  key  data  in other  spectral  windows  necessary  to  complement  the  radio  information  to  maximise  the scientific  outputs  of  these  projects. 
 This work is particularly useful for paving
the way to upcoming radio surveys that these new radio facilities will provide.

\section*{Acknowledgements}
We thank the staff of the GMRT that made these observations possible. GMRT is run by the National Centre for Radio Astrophysics of the Tata Institute of Fundamental Research. We also thank Claudia Mancuso and Anna Bonaldi for providing us their model predictions.
E.F.O thanks  the anonymous
referee for their comments on the manuscript. This work is based in part on observations made with the \textit{Spitzer Space Telescope}, which is operated by the Jet Propulsion Laboratory, California Institute of Technology under a contract with NASA.
We acknowledge support from the Italian Ministry of Foreign Affairs and International Cooperation (MAECI Grant Number ZA18GR02) and the South African Department of Science and Technology's National Research Foundation (DST-NRF Grant Number 113121) as part of the ISARP RADIOSKY2020 Joint Research Scheme.
\appendix
\section{POSTAGE STAMPS}
\begin{figure}
\centering
  \begin{tabular}{@{}cccc@{}}
    \includegraphics[width=.38\textwidth]{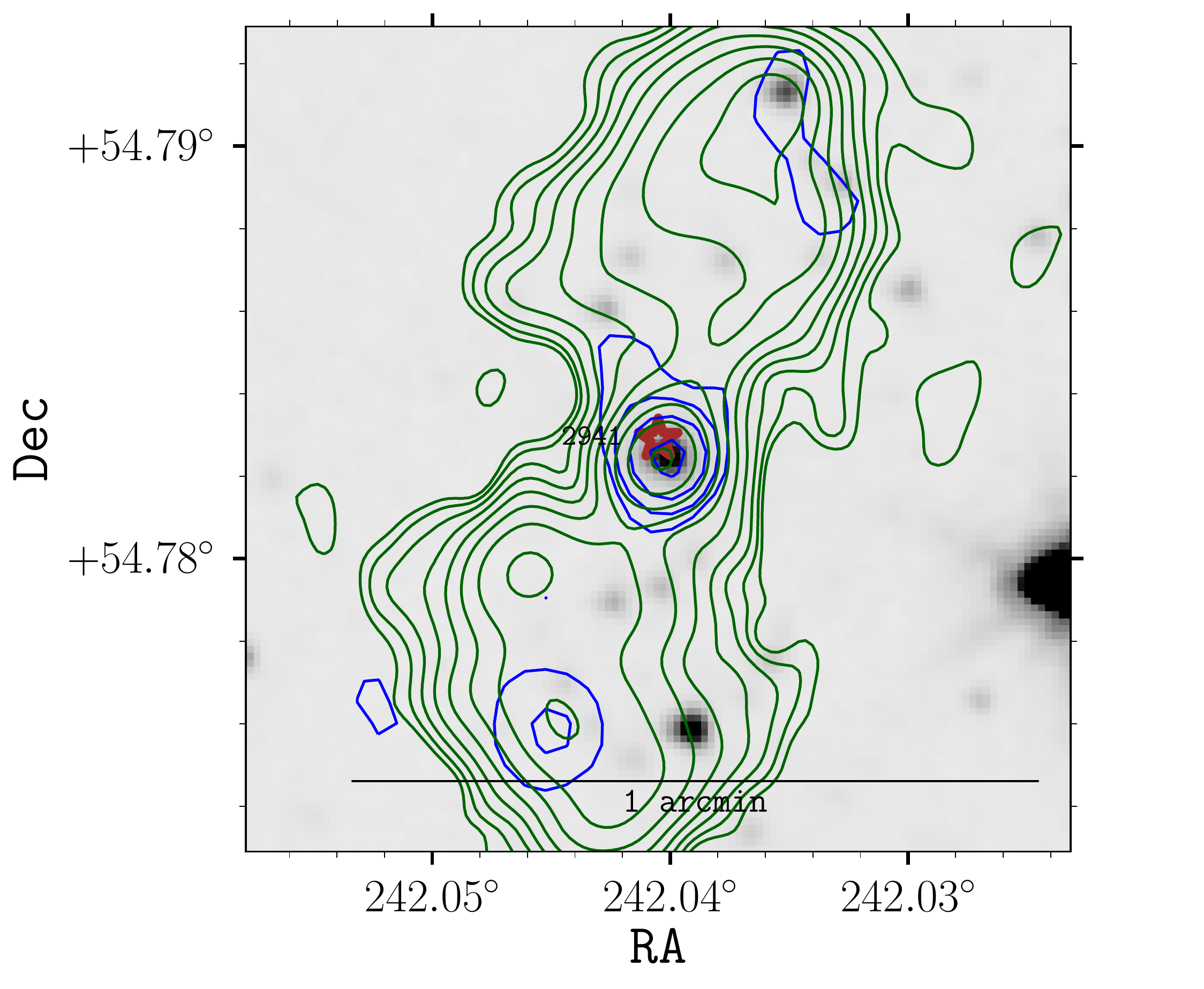} \\
    \includegraphics[width=.38\textwidth]{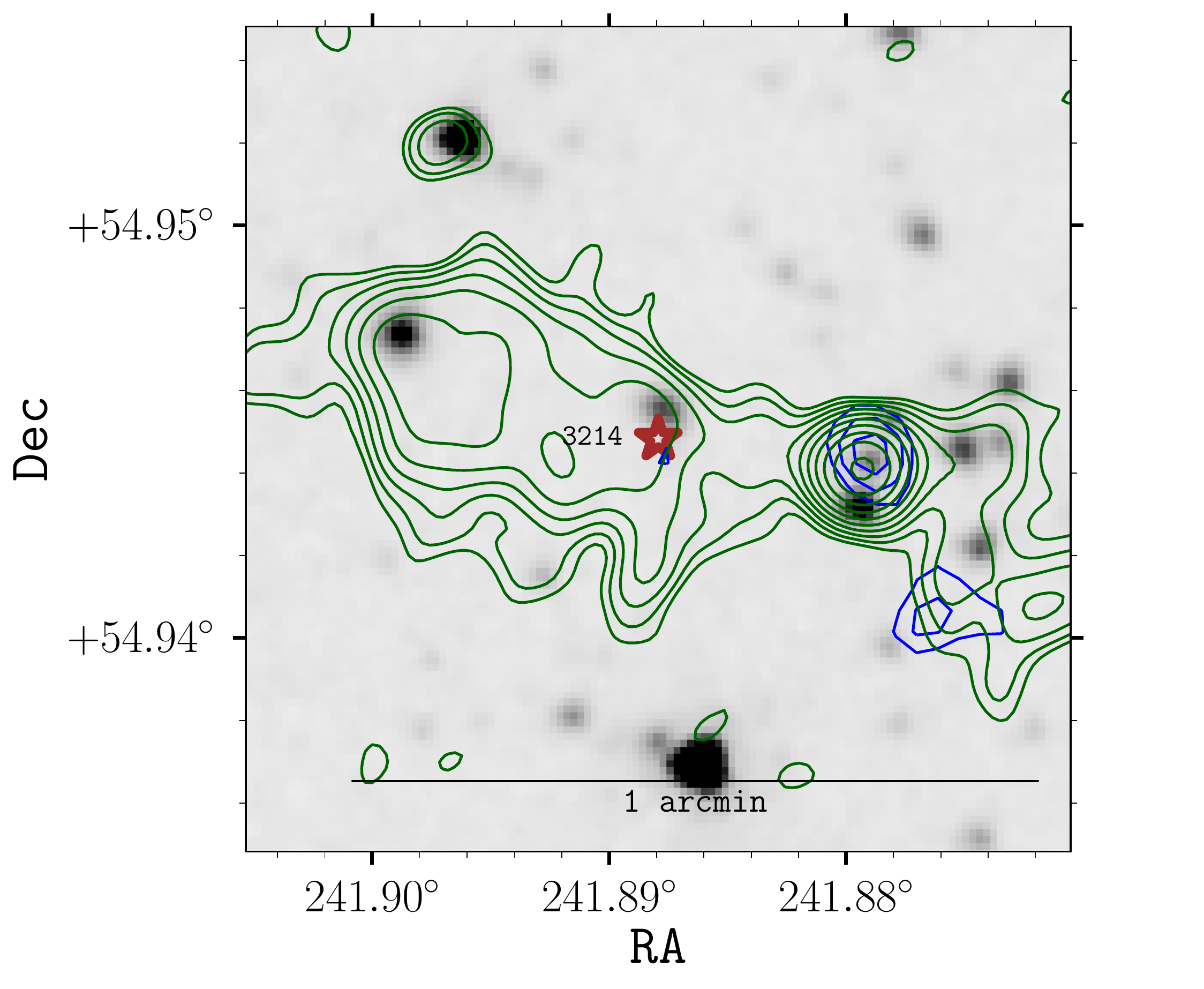} \\
    \includegraphics[width=.38\textwidth]{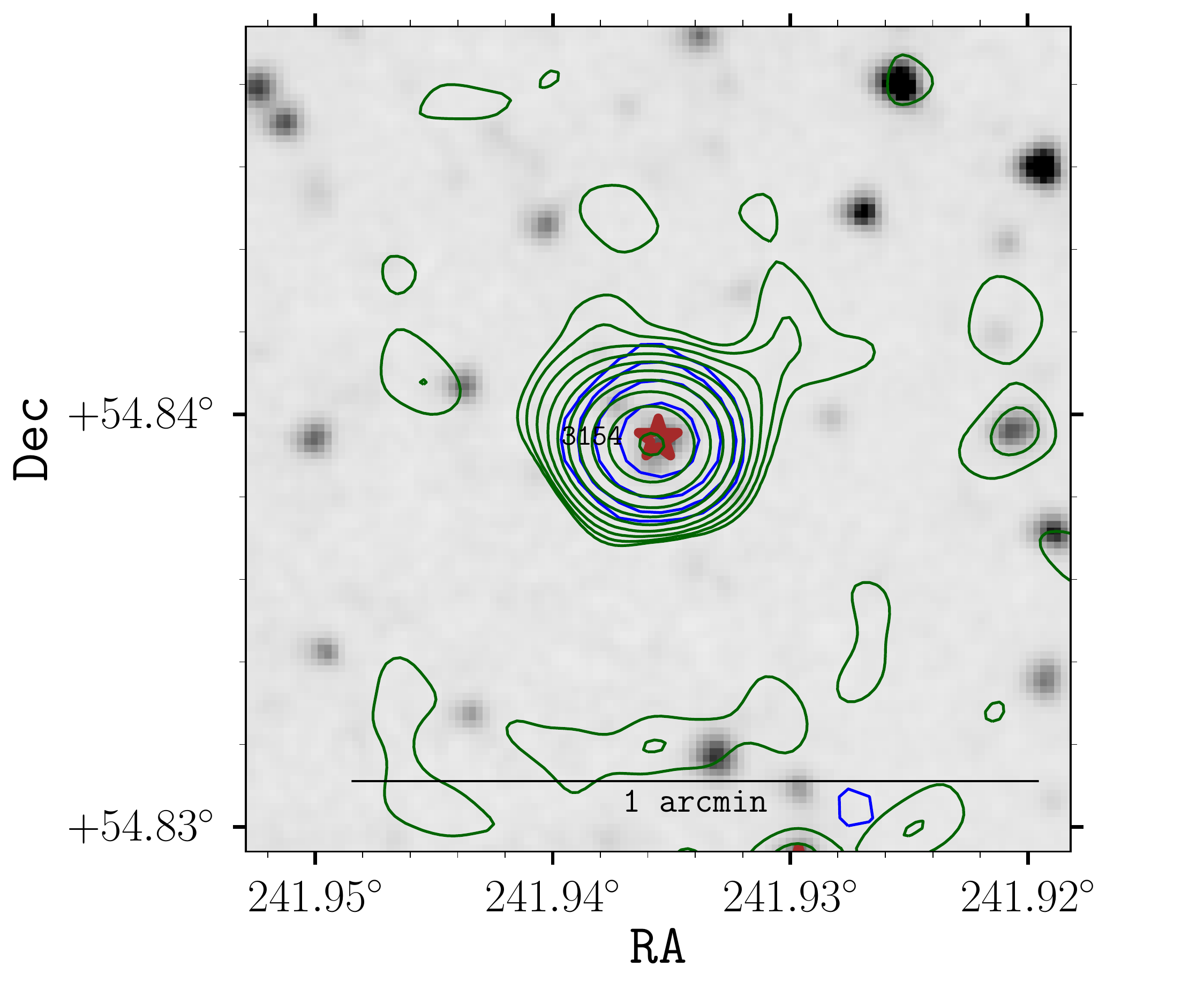} \\
    \includegraphics[width=.38\textwidth]{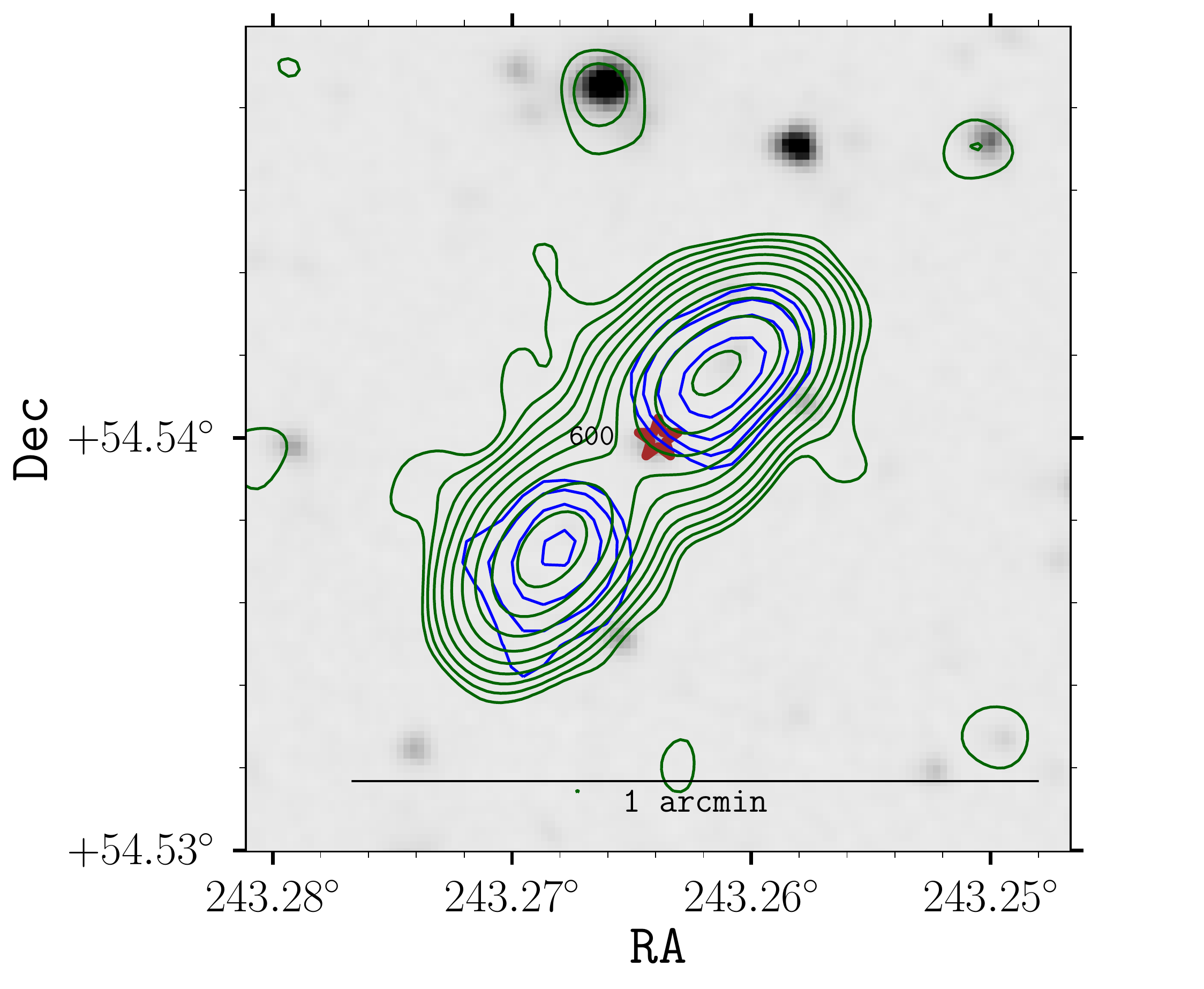} \\
  \end{tabular}
  \caption{Postage stamps showing examples of extended sources in the GMRT 610 MHz catalogue. The greyscale shows IRAC band 1 images. The green contours represents the GMRT 610 MHz whereas the blue contours represents VLA FIRST.}
\label{fig:postage_stamps}  
\end{figure}

Postage stamps of extended sources, including those merged into single sources, are shown in Figure~\ref{fig:postage_stamps}.
\bibliographystyle{mnras}
\bibliography{main} 
%
\bsp	
\label{lastpage}
\end{document}